\documentclass{aa}  

\usepackage{graphicx}
\usepackage{txfonts}

\newcommand{\Ms}{\ensuremath{M_{\odot}}}
\newcommand{\eg}{{\it e.g.}}

\newcommand{\ie}{{\it i.e.}}
\newcommand{\viz}{{\it viz.}}
\newcommand{\beq}{\begin{equation}}
\newcommand{\eeq}{\end{equation}}

\newcommand{\vg}{\ensuremath{v_g}}
\newcommand{\tg}{\ensuremath{\tau_g}}
\newcommand{\td}{\ensuremath{\tau_d}}
\newcommand{\tcr}{\ensuremath{\tau_{cr}}}

\newcommand{\kmps}{\ensuremath{\rm~km~s}^{-1}}

\begin{document}

   \title{The bound fraction of young star clusters}


   \author{Nina Brinkmann
          \inst{1}
          \and
          Sambaran Banerjee\inst{2,3}
          \and
          Bhawna Motwani\inst{4}
          \and
          Pavel Kroupa\inst{3,5}
          }
   \institute{Max-Planck-Institut f\"ur Radioastronomie, Auf dem Hügel 69,
              D-53121 Bonn, Germany\\
              \email{brinkmann@mpifr-bonn.mpg.de}
         \and
            Argelander-Institut f\"ur Astronomie, Auf dem Hügel 71,
            D-53121 Bonn, Germany\\
         \and
             Helmholz-Institut f\"ur Strahlen-und Kernphysik, Auf dem Hügel 71, D-53121 Bonn, Germany\\
         \and
            California Institute of Technology,
            1200 E. California Blvd., Pasadena, CA 91125, U.S.A.\\
         \and
            Charles University in Prague, Faculty of Mathematics and Physics, Astronomical Institute, V  Hole\v{s}ovi\v{c}k\'ach 2, CZ-180 00 Praha 8, Czech Republic
             }

   \date{Received:~~~~~~~~~~~~; Accepted:~~~~~~~~~~~~}

  \abstract
   {The residual gas within newly formed star clusters is expelled
through stellar feedback on timescales $\lesssim 1\ \mathrm{Myr}$.
The subsequent expansion of the cluster
results in an unbinding of a fraction of stars, before the
remaining cluster members can re-virialize and form a surviving cluster.}
   {We investigate the bound fraction after gas expulsion as a
function of initial cluster mass in stars $M_\mathrm{ecl}$ and gauge the
influence of primordial mass segregation, stellar evolution and the
tidal field at solar distance. We also assess the impact of the
star-formation efficiency $\varepsilon_\mathrm{SFE}$ and gas expulsion velocity $v_{g}$.}
   {We perform N-body simulations using Sverre Aarseth's {\tt NBODY7} code, starting with
compact clusters in their embedded phase and approximate the gas expulsion
by means of an exponentially depleting external gravitational field.
We follow the process of re-virialization through detailed monitoring of
different Lagrange radii over several Myr, examining initial half-mass radii
of $0.1 ~ \mathrm{pc}$, $0.3 ~ \mathrm{pc}$ and $0.5 ~ \mathrm{pc}$ and
$M_\mathrm{ecl}$ usually ranging from $5 \times 10^{3} ~ \mathrm{M}_{\odot}$ to $5 \times 10^{4} ~ \mathrm{M}_{\odot}$.}
   {The strong impact of the relation between the gas expulsion timescale
and the crossing time means that clusters with the same initial core
density can have very different bound fractions. The adopted $\varepsilon_\mathrm{SFE}=0.33$
in the cluster volume results in a distinct sensitivity to $v_{g}$ over a wide mass range, while a variation of
$\varepsilon_\mathrm{SFE}$ can make the cluster robust to the rapidly decreasing
external potential. We confirm that primordial mass segregation leads to a smaller
bound fraction, its influence possibly decreasing with mass. Stellar evolution has a
higher impact on lower mass clusters, but heating through dynamical friction could
expand the cluster to a similar extent. The examined clusters expand
well within their tidal radii and would survive gas expulsion even in a strong tidal field.}
   {}

   \keywords{galaxies: star clusters --
                methods: N-body simulations
               }

   \maketitle
%

\section{Introduction}\label{intro}

Star clusters form within the dense cores of giant molecular clouds \citep{LadaLada2003,Alves2012,Megeath2016}.
Within proto-cluster cloud cores, individual proto-stellar cores approach their hydrogen-burning
main sequences to form an infant star cluster, still embedded in residual gas.
Both observational and numerical studies \citep{AdamsFatuzzo1996} indicate that star formation process may be self-regulatory,
due to mechanical and radiation feedback from the proto-stars and O/B main-sequence stars,   
leading to a partial conversion of gas into stars. This means that all newborn clusters
begin their lives in a gas-embedded phase.   
The incomplete conversion from gas into stars can be quantified through the
star-formation efficiency ($\varepsilon_\mathrm{SFE}$)
\begin{equation}
 \varepsilon_\mathrm{SFE}=\frac{M_\mathrm{ecl}}{M_\mathrm{ecl}+M_\mathrm{gas}},
\label{eq:sfedef}
\end{equation}
where, $M_\mathrm{ecl}$ and $M_\mathrm{gas}$
describe the total mass in stars and gas respectively, right after star formation ceases
within the proto-cluster clump. This assumes star cluster formation
over a ``single'' starburst within the dense gas clump. A lack of age spread
in young massive star clusters in the Milky Way and the Magellanic Clouds 
support episodic formation of star clusters. Also, as demonstrated in \citet{BanerjeeKroupa2015},
a short-duration cluster assembly implies formation of either a monolithic
stellar distribution or close to monolithic one. In the latter case, a substructured
stellar distribution is formed, where the stellar overdensities (sub-clusters) are
located closely at birth (due to triggering) and merge into a single cluster in
$\lesssim1$ Myr.

Both, observational \citep{LadaLada2003,Megeath2016} and computational studies \citep{MM2012} suggest $\varepsilon_\mathrm{SFE}$ $\lesssim30$\%
within dense star-forming gas clumps, which are typically of parsec scale. Over molecular
clouds of $\gtrsim10$ pc, $\varepsilon_\mathrm{SFE}$ is only a few percent. The newly-formed cluster remains
gas embedded until a substantial portion of the gas is ionized by the UV radiation
from the O/B stars. Once ionized, the radiation couples efficiently
and pressurizes the gas over the entire proto-cluster, ultimately blowing it off the
cluster. The properties of radiation propagation in ultra-compact $\ion{H}{ii}$ (UC$\ion{H}{ii}$) regions
dictate a relatively brief phase of $\td\lesssim1$ Myr, until the residual gas is expelled
\citep{SambaranPavel2013}. Indeed, the existence of few-Myr old,
well-exposed massive starburst clusters,
\eg, $\approx1$ Myr-old NGC 3603, suggests that the gas-embedded phase lasts for
$\lesssim 1\ \mathrm{Myr}$. Such radiative gas expulsion would take place
at least with the sound speed in ionized hydrogen ($\approx10\kmps$),
and hence can be expected to happen promptly,
typically faster than or similar to the dynamical time or crossing time, $\tcr$, of
the embedded cluster \citep{SambaranPavel2013,LadaLada2003}.
The corresponding dilution of the potential well causes the cluster to expand in its
dynamical timescale. In this process, a fraction of the originally bound stars
becomes unbound and leaves the cluster, before the latter may reach a new equilibrium state
(\eg, \citealt{Adams2000,KAH2001,BoilyKroupa2002,Baumgardt2007,Pfalzner2013,SambaranPavel2013,SmithGoodwin2013}).

The timescale of gas expulsion, $\tau_\mathrm{g}$, and $\varepsilon_\mathrm{SFE}$
are the two key parameters that determine the evolution of a given embedded
cluster \citep{Lada1984,LadaLada2003,Adams2000}. For a collisionless system, gas removal
should unbind a cluster (or any gravitationally self-bound system)
if $\varepsilon_\mathrm{SFE}\leq 0.3$. The survivability of star
clusters, for $\varepsilon_\mathrm{SFE}\leq 0.3$, relies solely on
energy exchange among stellar orbits
over a dynamical timescale in the expanding cluster. This causes a fraction of the stars,
preferentially from the cluster's central region with highest stellar density,
to lose orbital energy and eventually fall back to form a remaining bound
cluster which is in dynamical equilibrium, \ie, ``re-virialized''
\citep{SambaranPavel2013,KAH2001}. A longer $\tg$, therefore,
allows for a longer time for violent relaxation, resulting in 
a larger remnant cluster. This mechanism plays a crucial role
for a cluster's survival, since both
observations and theoretical studies support clump
$\varepsilon_\mathrm{SFE}$ of $\lesssim0.3$ \citep{LadaLada2003,MM2012,Megeath2016}.

In this study, we consider initially massive, compact clusters and investigate
the mass bound fraction, \ie, the mass fraction of stars that remain gravitationally bound
after the gas dispersal, to form a surviving cluster. While this is a basic question which
has already been explored in the literature \citep{Lada1984}, its importance is renovated as an increasing
number of young massive and open clusters, with a variety of dynamical conditions,
are being observed in detail. This issue is directly connected to the question of   
the contribution of clustered star formation to the field star population in a galaxy.
Despite several notable contributions over the past decade, a systematic study
of the bound fraction of star clusters with realistic conditions is still pending.
Assuming a canonical initial mass function (IMF), there are four primary parameters
that control an embedded cluster's bound fraction after its re-virialization, \viz,
its initial mass, $M_\mathrm{ecl}$, its initial size (half-mass radius), $r_\mathrm{h}$,
$\varepsilon_\mathrm{SFE}$, and gas dispersal timescale, $\tg$.
%

\section{Initial conditions and gas expulsion}\label{models}

The starting point for all our simulations is a star cluster in its gas-embedded phase.
The total initial stellar mass, $M_\mathrm{ecl}$, is compounded by stars following the
canonical optimal IMF \citep{KroupaWeidner2013}, whose spatial and velocity distributions
are described by a Plummer sphere in dynamical equilibrium \citep{HH2003,Plummer1911}.
Initially dense spherical Plummer conditions are a good approximation
given the structure of well observed very young clusters \citep{KAH2001,SambaranPavel2013}.
Initially hierarchical sub-clump configurations are limited to
be near-monolithic systems formed in a single-burst, given its youth \citep{BanerjeeKroupa2015}.
The residual gas is mimicked by a time-dependent external potential of a spherically-symmetric mass distribution of total mass
$M_\mathrm{gas}$ following the same Plummer radial
density profile as the stars, enabling us to effectively adopt
an $\varepsilon_\mathrm{SFE}$ as defined by Eqn.~\ref{eq:sfedef}.
 The essential dynamical effect of
gas dispersal is mimicked
by depleting the total gas mass $M_\mathrm{gas}$ (and hence
its potential), exponentially: 

\begin{equation}
 M_\mathrm{gas}\left(t\right)=
M_\mathrm{gas}\left(0\right)  ~ \exp
\left( - \frac{t-\tau_\mathrm{d}}{\tau_\mathrm{g}} \right) ~ , ~ t \geq \tau_\mathrm{d},
\label{eq:mgas}
\end{equation}

thus allowing the embedded cluster to evolve secularly for time $\tau_\mathrm{d}$,
before $M_\mathrm{gas}$ decreases in a timescale $\tau_\mathrm{g}$. This timescale,
which is given by

\begin{equation}
\tau_\mathrm{g}= \frac{r_\mathrm{h}\left(0\right)}{v_\mathrm{g}} ~ ,
\label{eq:tg}
\end{equation}

depends on the initial half-mass radius (HMR), $r_\mathrm{h}\left(0\right)$, of the stellar cluster and the gas,
and the average radial velocity of gas expulsion, $v_\mathrm{g}$.
The latter is taken to be $v_\mathrm{g} \approx 10 ~ \mathrm{km ~ s^{-1}}$, the
sound speed in an $\ion{H}{ii}$ region (\eg, \citealt{KAH2001,SambaranPavel2013}).
This simplifying assumption neglects the possible radiation-pressure dominated (RPD)
initial phase of gas expansion, during which the gas blow-out velocity might exceed
the speed of sound (which is essential for massive clusters with large escape velocities),
as shown by \citet{KrumholzMatzner2009},
thereby denoting our choice of $\tau_\mathrm{g}$ as an upper limit.

The time until the commencement of gas expulsion is estimated to be $\tau_\mathrm{d}\approx0.6 ~ \mathrm{Myr}$
(see \citealt{KAH2001,SambaranPavel2013} and references therein),
by considering the lifetimes of ultra-compact
($\approx0.1 ~ \mathrm{pc}$) $\ion{H}{ii}$ regions of up to $\approx0.1 ~ \mathrm{Myr}$
and scaling them to the radii investigated in our simulations \citep{Sambaran2014}.
During the time until termination of star formation, several generations
of stars can condense out of the gas \citep{WuchterlTscharnuter2003} in a real cluster
and orbit the potential for several crossing times, thereby nearing equilibrium (see also \citealt{Kroupa2005}).

Admittedly, the process of gas dispersal, especially from massive clusters,
is still a poorly understood process and the quantities, $\vg$($\tg$) and $\td$, as
adopted above to describe this process have to be treated essentially as model parameters, with
their plausible representative values chosen based on the present knowledge.
If not stated otherwise, we treat the models computed here as
isolated clusters with no primordial mass segregation;
stellar evolution is taken into account and the initial HMR is $0.3 ~ \mathrm{pc}$ and
$\varepsilon_\mathrm{SFE}=0.33$.
The above mentioned values for $v_\mathrm{g}$ and $\tau_\mathrm{d}$ are used for
the time evolution of the gas potential.

\section{Bound fraction and its dependence on cluster parameters}\label{bfrac}

The effect of gas expulsion from an embedded cluster
is an interplay between the rapid (in timescale
comparable to the dynamical time of the cluster) dilution of the gas potential
that tends to unbind the cluster and orbital energy exchange among
the stars (in a rapidly-changing gravitational field) resulting in ``violent
relaxation'' that helps a fraction of the expanding stellar population
to dissipate energy and fall back and retain a (lower-mass) bound cluster
(see \citealt{2015arXiv151203074B} for a discussion). A self-consistent and
accurate treatment of this violent relaxation is crucial
for a reliable estimate of the bound fraction left after the gas removal.
This is why we need to resort to direct N-body simulations in
this work.
  
The dynamical relaxation of a self-gravitating many body system is most realistically
calculated using star-by-star direct N-body integration. In this study, we use the 
state-of-the-art direct N-body
code, {\tt NBODY7} (formerly {\tt NBODY6}; \citealt{aseth2003,aseth2012}),
to compute the evolution of star clusters before, during and
following gas expulsion.
In addition to tracking the individual stars' orbits using the highly-accurate
fourth-order Hermite scheme and dealing with the diverging gravitational forces, during, \eg, close
encounters and hard binaries,
through two-body and many-body regularizations, {\tt NBODY7} also includes
well-tested analytical stellar and binary evolution recipes
\viz, the {\tt SSE} and the {\tt BSE} schemes \citep{hur2000,hur2002}, 
that is employed simultaneously with the dynamical
integration. The above stellar-evolutionary schemes include
mass loss due to stellar winds and supernovae and stellar masses and
other parameters are updated accordingly during the N-body integration.
Over the early ages considered in this study ($\lesssim 20$ Myr),
the stellar mass loss happens predominantly due to very fast winds from
O-type stars ($\sim10^3 \kmps$) and supernova ejecta ($\sim10^4 \kmps$),
which would escape from the cluster immediately. Hence, in these calculations,
any stellar mass loss is simply removed from the system,
contributing to the dilution of the cluster's gravitational potential
accordingly.
Furthermore, mergers among stars and remnant formation
are treated through analytic recipes. Most importantly, no softening of gravitational forces
is employed at any stage, making sure that the energetics of general
two-body interactions and close encounters,
that plays a key role in the dynamical relaxation of the cluster, is accurately
computed.
This numerical code, therefore, naturally and best suits the purpose of this
study.   

A straightforward way to calculate the bound fraction of a computed model cluster,
as a result of its early gas expulsion,
is to follow its Lagrange radii over time.
In this study, we use Lagrange radii at 2\% intervals of mass fraction, 
reaching from 2\% to 99\%. This allows for the tracking of the different
layers of an evolving cluster fairly closely and hence the calculation of the bound fraction with reasonable
accuracy, for a particular simulation (see, e.g., Fig. \ref{CompSFE} below). Here,
the bound fraction after gas expulsion is determined by the
outermost Lagrange radius displaying a reversal of its expansion,
though we also take the slope of the radii in comparison to the
bound inner layers into account. The re-collapse of a particular Lagrange radius
can be easily identified by eye estimation (c.f. Fig. \ref{CompSFE}; except perhaps
for the innermost few). That way the `inherent' bound fraction,
as a result of the loss of stars due to gas expulsion
(plus any stellar-evolutionary mass loss), is estimated.  
The fact that one does get consistent results, as shown below,
justifies this approach.
 
Note that this approach does not strictly take into account the stars'
gravitational boundedness to the cluster. This is because it can, in principle,
include a small fraction of stars whose velocities are higher than the
escape velocity and which are on their way of escaping the system. 
We plot the bound fraction as a function of $M_\mathrm{ecl}$ for most considered parameters.
The functions tracing the bound fractions are basic
$\arctan ~ \mathrm{functions}$ which are fitted to the data to provide visual aid only. 

Notably, the relative bound mass after $20\ \mathrm{Myr}$, as a function of $\varepsilon_\mathrm{SFE}$
and initial cluster density, is studied by \citet{Pfalzner2013},
whereas our results give an upper limit on the bound fraction for a given initial condition,
without focusing on a certain point in time.
Their bound fractions after $20\ \mathrm{Myr}$ would consequently be
smaller than those determined with our method. 
In addition to the overall bound fraction of a computed cluster,
we also note the re-virialization time for 10\%, 20\%, etc. Lagrange radii.
The realized simulations are intended to give an impression of the impact of
various physical parameters on the bound fraction, providing a base for further studies.
The error bars in bound fractions, for most plots here, are derived using the
2\% error in determining the bound fraction in an individual simulation;
they are sometimes larger due to fluctuations in the Lagrange radii, especially for lower mass clusters.
When multiple runs are available for a particular set of parameters,
we use the mean values and standard deviation.

\subsection{Half-mass radii and re-virialization times}\label{tvir}

\begin{figure}
\resizebox{\hsize}{!}{\includegraphics[angle=-90]{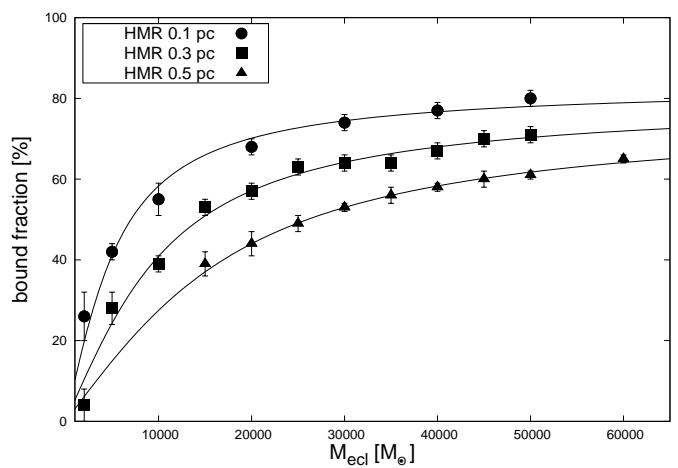}}
\caption{The bound fraction for initial HMR of $0.1 ~ \mathrm{pc}$,
$0.3 ~ \mathrm{pc}$ and $0.5 ~ \mathrm{pc}$ as a function of initial cluster mass.
The data points for $0.1 ~ \mathrm{pc}$ and $0.3 ~ \mathrm{pc}$ correspond to one run per mass,
for $0.5 ~ \mathrm{pc}$ 3-6 are averaged.}
\label{HMRdiff}
\end{figure}

\begin{figure}
\resizebox{\hsize}{!}{\includegraphics[angle=-90]{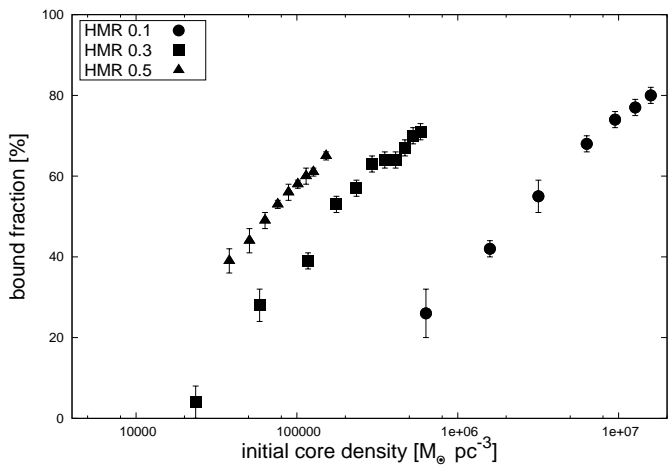}}
\caption{The bound fraction as a function of initial core density
for initial HMR of $0.1 ~ \mathrm{pc}$, $0.3 ~ \mathrm{pc}$ and $0.5 ~ \mathrm{pc}$.}
\label{BFvsRho}
\end{figure}

To examine the influence of certain key cluster parameters on the bound fraction after gas expulsion,
we consider massive and initially very compact clusters with a HMR of $\lesssim 0.3 ~ \mathrm{pc}$.
Such compact initial conditions are consistent with the sizes of the observed proto-cluster cores
and filament junctions in molecular clouds \citep{Malinen2012,Andre2014}, where massive clusters are likely to form. 
In order to make estimations for differently sized clusters, we typically take initial HMRs of
$ r_\mathrm{h} = 0.1 ~ \mathrm{pc}$, $0.3 ~ \mathrm{pc}$ and $0.5 ~ \mathrm{pc}$
and plot their bound fraction as a function of initial cluster mass. This is shown in Fig.\ref{HMRdiff}.
Here, the $\arctan ~ \mathrm{funcions}$ used for fitting suggest a theoretical maximum
bound fraction of roughly 83\% for the compact $0.1 ~ \mathrm{pc}$, and 77\% for the $0.5 ~ \mathrm{pc}$ clusters.
Over the whole mass range, the initial HMR seems to have a distinct effect on the bound fraction.
This dependency also shows up when plotting the bound fraction as a function of initial core density (see \citealt{Kroupa2008initial}, particularly pages 28 and 31), 
\begin{equation*}
\rho_\mathrm{core} = \frac{M_\mathrm{core}}{\frac{4}{3} \times \pi \times r_\mathrm{core}^{3}} \approx 0.32 ~ \frac{M_\mathrm{ecl}}{r_\mathrm{h}^{3}} ~ , 
\end{equation*}
as shown in Fig.~\ref{BFvsRho}: the same initial core density can lead to very different bound fractions,
depending on the initial HMR.

\begin{figure}
\resizebox{\hsize}{!}{\includegraphics[angle=-90]{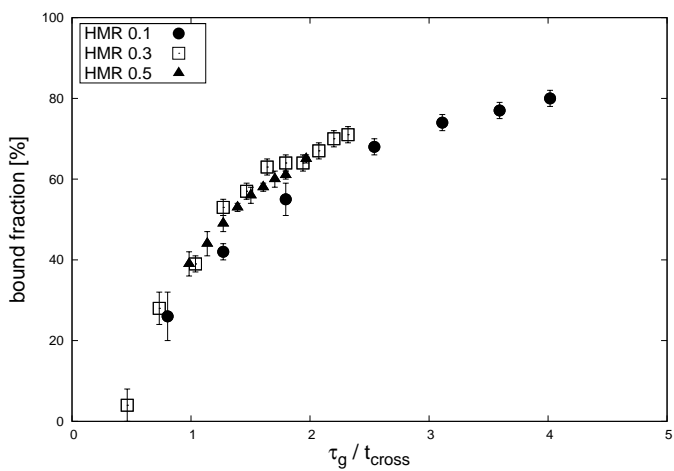}}
\caption{The bound fraction as a function of the ratio between gas
expulsion timescale and crossing time. A certain bound fraction may be traced
back to a specific value of $\tau_\mathrm{g} / t_\mathrm{cross}$.}
\label{BFvsTRelation}
\end{figure}

This is mostly owed to the influence of the HMR on the important relation
$\tau_\mathrm{g} / t_\mathrm{cross}$ \citep{Baumgardt2007}.
With $\varepsilon_\mathrm{SFE}$ and gas expulsion velocity $v_\mathrm{g}$ both set
and the definitions given in \citet{Kroupa2008initial}, we can estimate
\begin{equation*}
 \frac{\tau_\mathrm{g}}{t_\mathrm{cross}} \propto
\sqrt{\frac{M_\mathrm{ecl}}{r_\mathrm{h}}}\qquad \text{respectively}\qquad\propto r_\mathrm{h}
\end{equation*}
for a fixed density: although a smaller cluster can re-virialize faster,
the timescale for gas expulsion decreases even quicker,
deteriorating the above relation and reducing the bound fraction.
This is emphasized when plotting (Fig. \ref{BFvsTRelation}) the bound fraction against
\begin{equation*}
 \frac{\tau_\mathrm{g}}{t_\mathrm{cross}} =
\sqrt{\frac{G \times M_\mathrm{ecl}}{\varepsilon_\mathrm{SFE} \times r_\mathrm{h}}} ~ \left(2\times v_{g} \right)^{-1} ~ ,
\end{equation*}
$G$ being the gravitational constant.
A specific value of $\tau_\mathrm{g} / t_\mathrm{cross}$ seems to result in a corresponding bound fraction,
a stronger correlation than observed for the other parameters.

\begin{figure}
\resizebox{\hsize}{!}{\includegraphics[angle=-90]{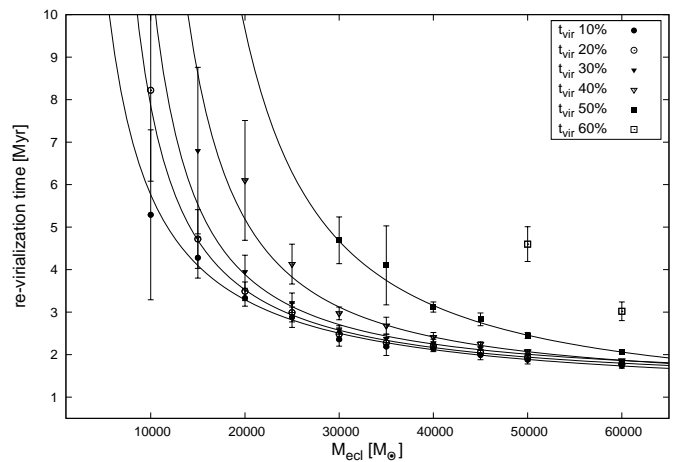}}
\caption{Re-virialization times for clusters with an initial HMR of
$0.5 ~ \mathrm{pc}$ and $M_\mathrm{ecl}$ of $10^{4} ~ \mathrm{M}_{\odot}$ to $6\times10^{4} ~ \mathrm{M}_{\odot}$.
The values are averaged over 3-6 simulations and include the
$\tau_\mathrm{d}\approx 0.6 ~ \mathrm{Myr}$ equilibrium (embedded) phase of the cluster.}
\label{TvirHMR05}
\end{figure}

The re-virialization time, here defined as the moment a given Lagrange radius no longer shrinks,
is more difficult to determine. Specifically in the lower mass clusters,
we find  ambiguity due to the effects of stellar evolution setting in and `overlaying' the contraction of
the Lagrange radii. The inner radii also pose difficulties caused by fluctuations due to
strong stellar encounters. The re-virialization times plotted in Fig.~\ref{TvirHMR05},
therefore, give only a rough estimate and are only adoptable for an initial
HMR of $0.5 ~ \mathrm{pc}$. The more compact $0.1 ~ \mathrm{pc}$ and $0.3 ~ \mathrm{pc}$ clusters
show considerably shorter re-virialization times ($< 1 ~ \mathrm{Myr}$ for most Lagrange radii for
the HMR $0.1 ~ \mathrm{pc}$ clusters). 

\subsection{Gas expulsion velocity}\label{gasvel}

\begin{figure}
\resizebox{\hsize}{!}{\includegraphics[angle=-90]{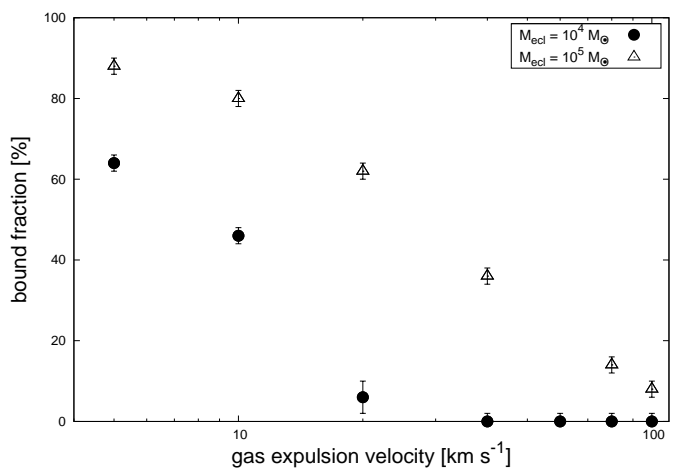}}
\caption{Influence of the gas expulsion velocity on the bound fraction
for initially unsegregated, isolated clusters with stellar masses of
$10^{4} ~ \mathrm{M}_{\odot}$ and $10^{5} ~ \mathrm{M}_{\odot}$,
considering a $\varepsilon_\mathrm{SFE}$ of $0.33$ and an initial
HMR of 0.3 pc, one simulation per data point.}
\label{PlotVgas_Mass}
\end{figure}

The impact of the gas expulsion velocity, $\vg$, on the bound fraction is examined for two clusters
with initial stellar component masses $M_\mathrm{ecl}$ of $10^{4} ~ \mathrm{M}_{\odot}$ and
$10^{5} ~ \mathrm{M}_{\odot}$,
by varying $v_\mathrm{g}$ from $5 ~ \mathrm{km~s^{-1}}$ to $100 ~ \mathrm{km~s^{-1}}$.
The high sensitivity of the clusters' survival to the changing velocity (see Sec.~\ref{intro})
is illustrated in Fig.~\ref{PlotVgas_Mass}. Infant mass loss differs strongly
over a comparatively short range of $\vg$, suggesting that a
$10^{4} ~ \mathrm{M}_{\odot}$ cluster, with $\varepsilon_\mathrm{SFE}=0.33$, would not survive
the gas expulsion phase, should the average gas velocity exceed $30 ~ \mathrm{km~s^{-1}}$.

\subsection{Varying star-formation efficiency}\label{VarSFE}

With the notable impact of the gas removal velocity on the bound fraction
as shown in the previous section, the variation of $\varepsilon_\mathrm{SFE}$ seems to be the
only way to produce clusters that survive their gas expulsion phase
when considering higher gas velocities, the often-assumed instantaneous gas removal
being the extreme case. The rest of the parameters, that are examined in our simulations, 
do not seem to have comparable effects on the infant weight loss.

The effect of variation of $\varepsilon_\mathrm{SFE}$, to reduce infant weight loss
or even to prevent the complete dissolution of a cluster of intermediate mass, is summarized
in Fig.~\ref{PlotVgas_SFE}, where we look upon unsegregated
$10^{4} ~ \mathrm{M}_{\odot}$ clusters. Progressing towards higher velocities,
while $\varepsilon_\mathrm{SFE} = 0.33$ will not result in a surviving cluster,
the bound fraction reaches a plateau for
$\varepsilon_\mathrm{SFE}= 0.50$ and $\varepsilon_\mathrm{SFE}=0.66$, rendering a cluster tolerant
against any gas loss. Simulations of early cluster evolution (\eg, \citealt{Pfalzner2013}) with
instantaneous gas removal, therefore, necessitates a high $\varepsilon_\mathrm{SFE}$,
with possible consequences for other cluster parameters.

\begin{figure*}
\centering
\includegraphics[width=4cm, angle=-90]{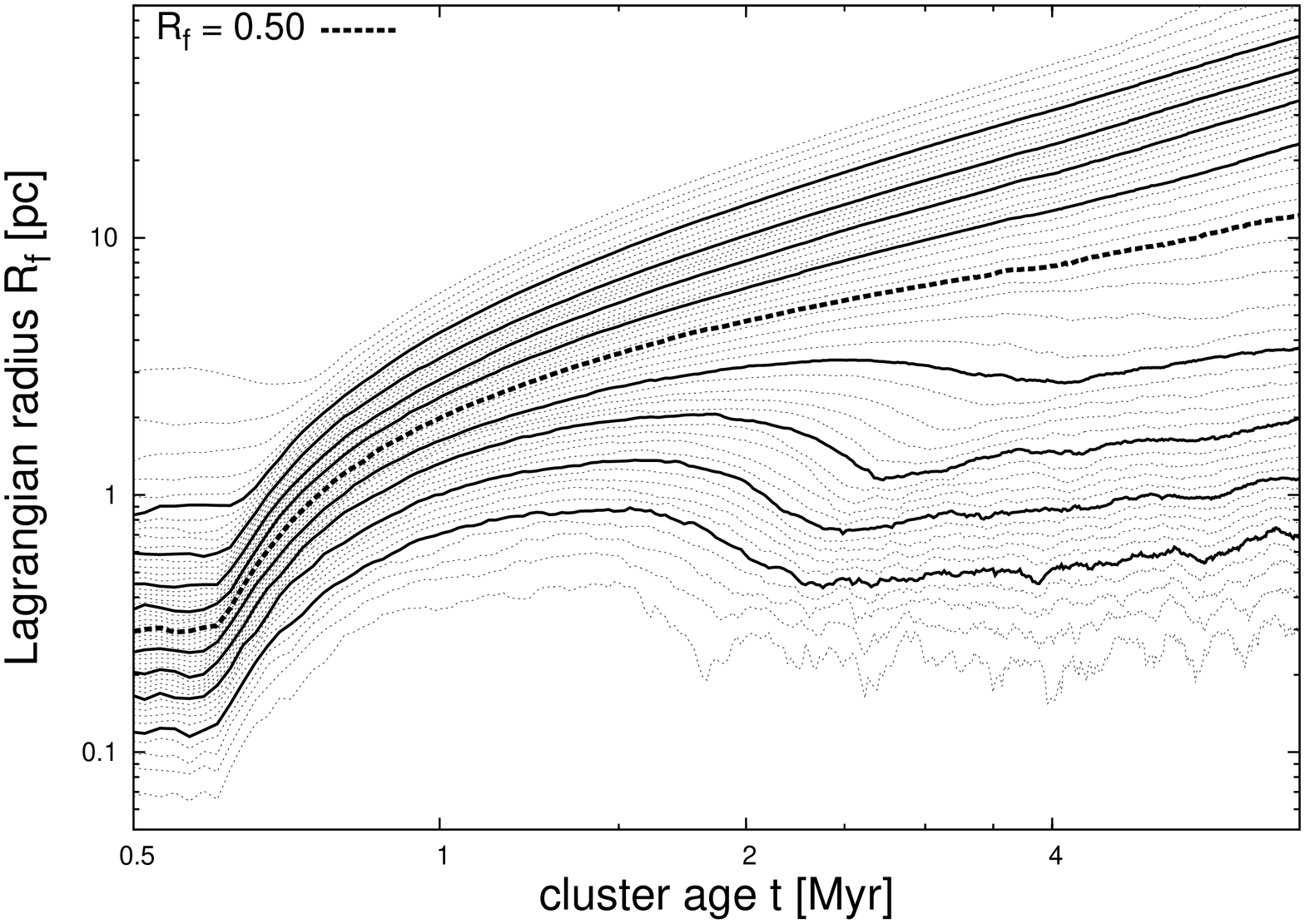}
\includegraphics[width=4cm, angle=-90]{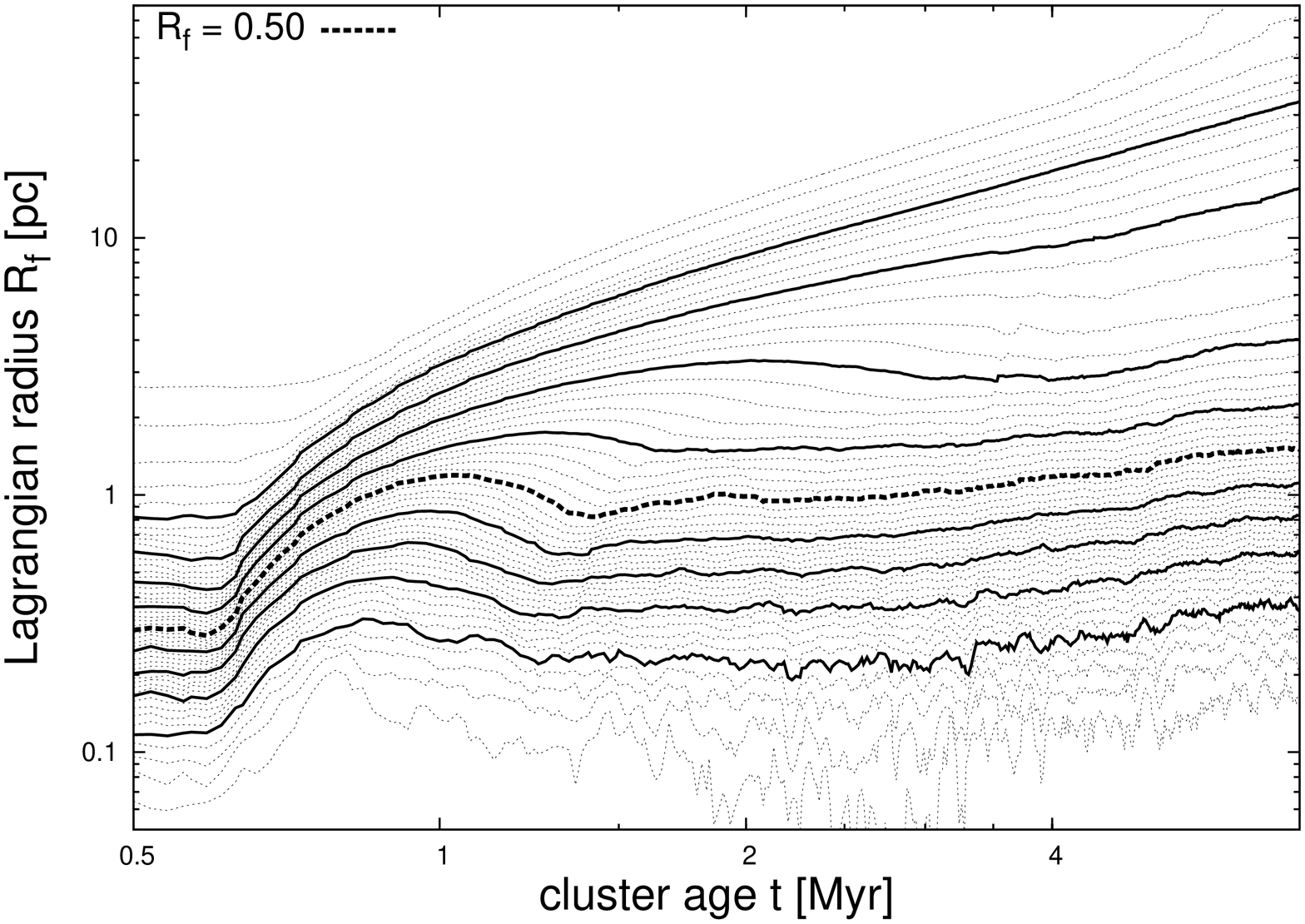}
\includegraphics[width=4cm, angle=-90]{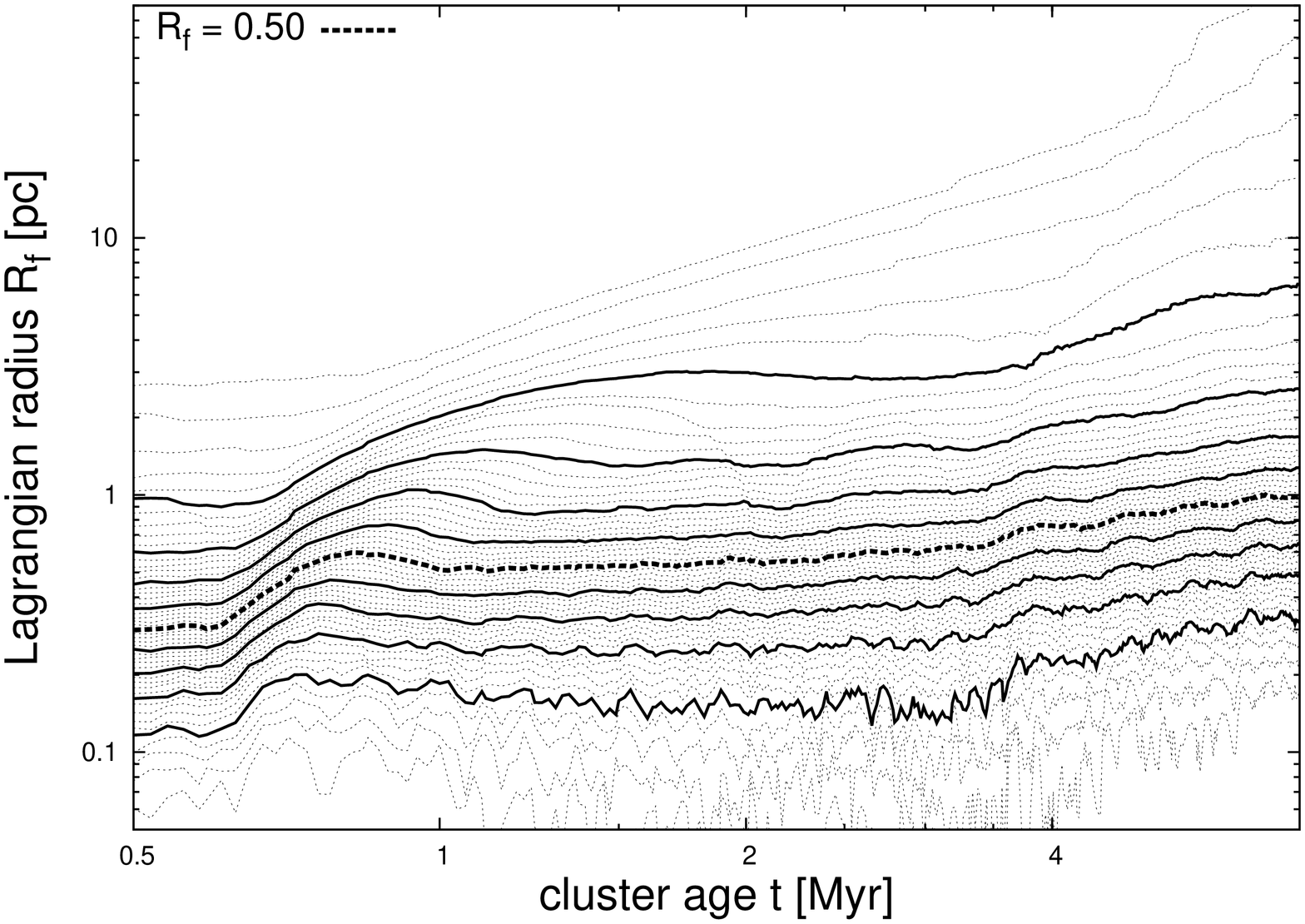}
\caption{Evolution over $7 ~ \mathrm{Myr}$ for $10^{4} ~ \mathrm{M}_{\odot}$ clusters with an initial
HMR of $0.3 ~ \mathrm{pc}$ for $\varepsilon_\mathrm{SFE}= 0.33$ (left panel),
$\varepsilon_\mathrm{SFE}= 0.50$ (middle panel) and $\varepsilon_\mathrm{SFE}= 0.66$ (right panel).
Pictured are 50 Lagrange radii, reaching from 2\% to 99\% in 2\% intervals.
As a visual aid, the prominent dashed curve refers to the HMR, the other prominent curves mark 10\% steps.}
\label{CompSFE}
\end{figure*}

\begin{figure}
\resizebox{\hsize}{!}{\includegraphics[angle=-90]{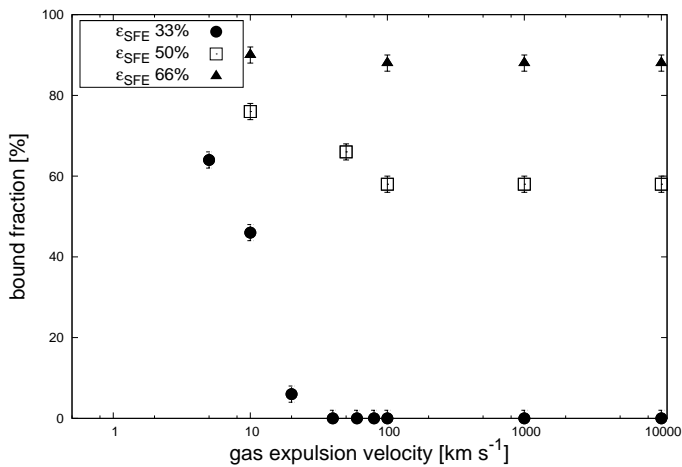}}
\caption{The bound fraction as a function of gas removal velocity for different $\varepsilon_\mathrm{SFE}$, here for clusters with an initial stellar mass of $10^{4} ~ \mathrm{M}_{\odot}$ and a HMR of $0.3\ \mathrm{pc}$. Approximating the gas removal to be instantaneous might make the assumption of an unusually high $\varepsilon_\mathrm{SFE}$ necessary, as otherwise simulations might not have any surviving cluster after gas expulsion. Every data point corresponds to a single simulation.}
\label{PlotVgas_SFE}
\end{figure}

A higher $\varepsilon_\mathrm{SFE}$ will result in a larger number of massive stars,
and hence more mass as massive stellar members,
compared to the primordial gas mass, for a given initial mass of the
cluster-forming gas clump.
This would, \eg, influence the clusters reaction to the effects of stellar evolution,
once the first SN occurs or result in a smaller re-virialization time (as the disturbance
caused by the rapid potential drop is smaller). This is demonstrated in Fig.~\ref{CompSFE}.
Additionally, the radial outgoing velocities of the unbound outer layers also
seem to be strongly affected by a change of $\varepsilon_\mathrm{SFE}$, \ie,
the radial velocities are smaller for larger $\varepsilon_\mathrm{SFE}$.

\subsection{Primordial mass segregation}\label{SEG}

Observations suggest that massive stars are preferentially located in the inner parts of star clusters,
both for young, exposed and embedded ones. Mass segregation is, of course, expected in dynamically-evolved
clusters as a consequence of two-body relaxation leading toward equipartition of energy
among stellar populations with different masses \citep{Meylan2000}.
However, mass segregation is also observed in a number of young, not yet relaxed systems,
suggesting primordial mass segregation \citep{deGrijsGilmore2002,
LittlefairNaylor2003,BontempsMotte2010} in them. Alternatively, diminished two-body relaxation
times in initially sub-structured stellar configurations may offer another explanation
for mass segregation in young clusters \citep{McMillanVesperini2007}.

The influence of primordial mass segregation on the bound fraction
is examined here by a series of simulations with
$M_\mathrm{ecl}$ ranging from $2\times10^{3} ~ \mathrm{M}_{\odot}$ 
to $5\times10^{4} ~ \mathrm{M}_{\odot}$. In these computed models, the
initial mass segregation is implemented using the method of \citet{bgetl2008}.
For lower to intermediate mass clusters, the bound fraction changes
by around 7\% to 10\% (even more for the lowest regarded mass), and by $\lesssim 3\ \mathrm{\%}$ for the high masses.

\begin{figure}
\resizebox{\hsize}{!}{\includegraphics[angle=-90]{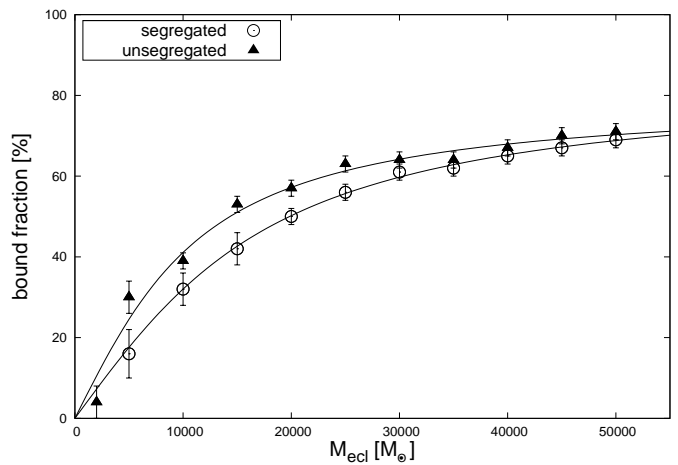}}
\caption{The influence of primordial mass segregation on the bound fraction as a function of initial cluster mass $M_\mathrm{ecl}$, based on one simulation per data point. An impact in lower mass clusters is observable, but the effect seems to lessen with increasing mass. This may be due to segregation in the heavy clusters in the equilibrium phase before gas expulsion.}
\label{SEG_Ser100Ser300}
\end{figure}

\begin{figure*}
\centering
\includegraphics[width=6cm, angle=-90]{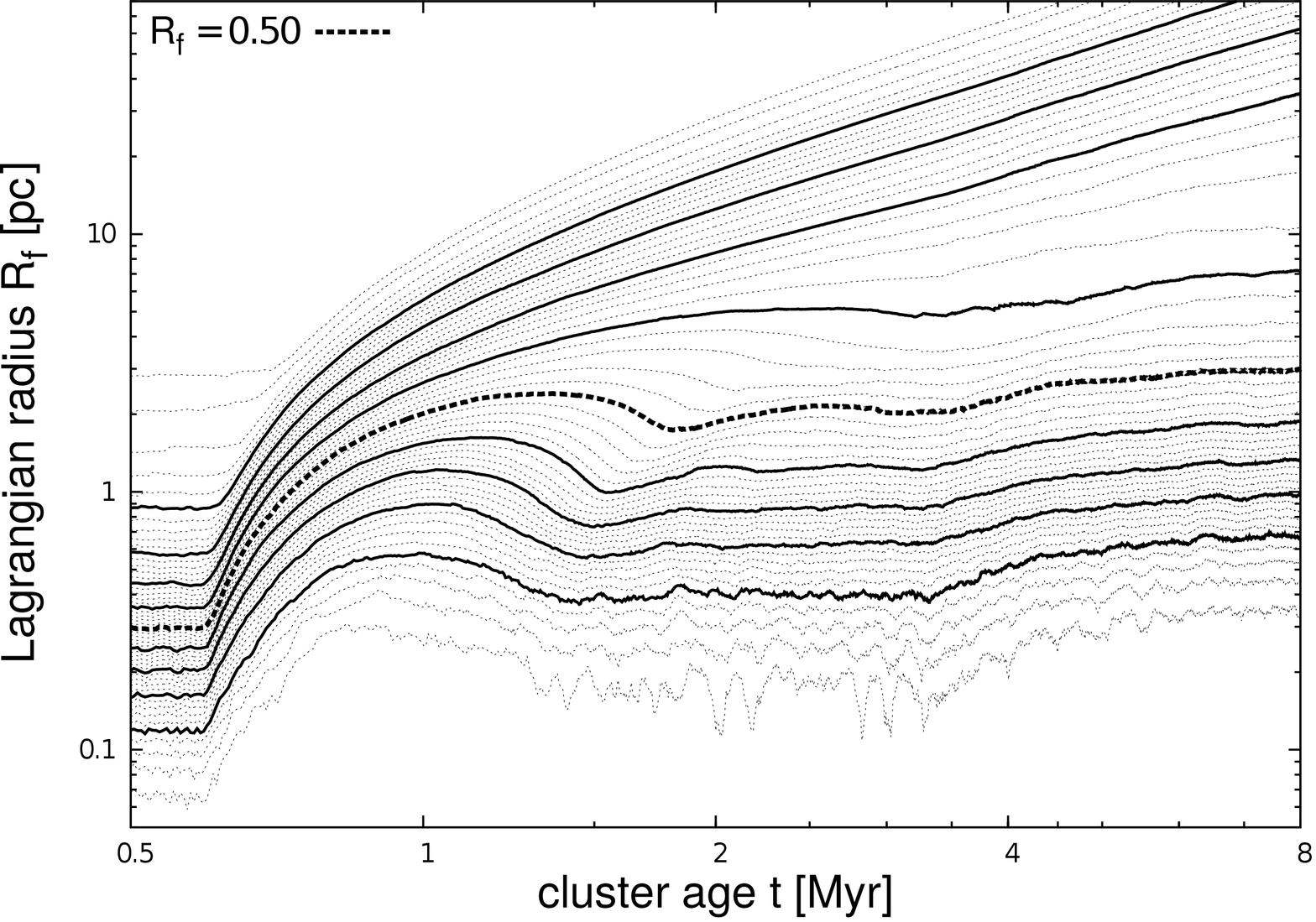}
\includegraphics[width=6cm, angle=-90]{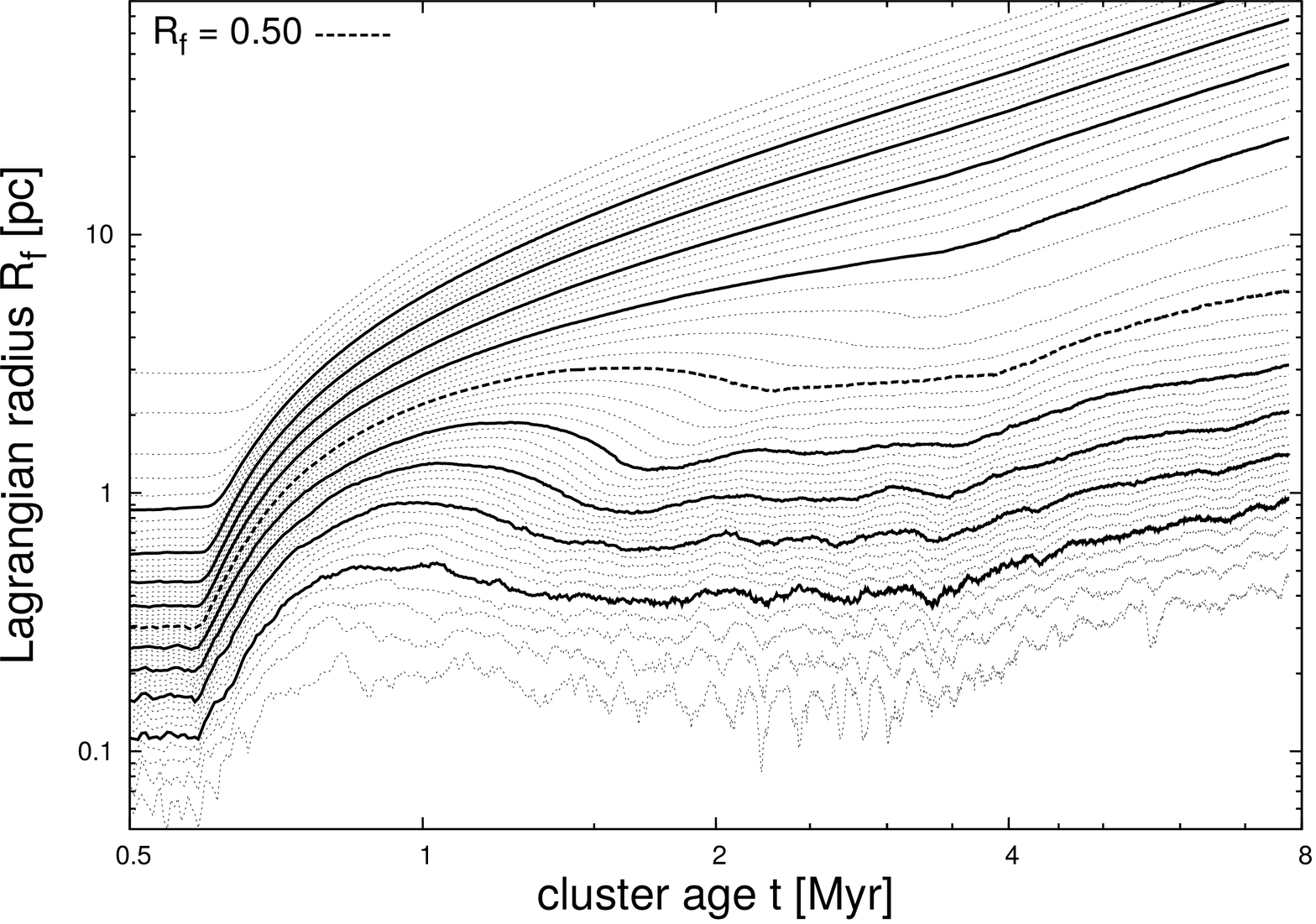}
\caption{Effect of primordial mass segregation. Tracking the first 8 Myr (respectively 7.4 Myr after the beginning of gas expulsion) for two $2.5\times10^{4} ~ \mathrm{M}_{\odot}$ clusters, without (left) and with (right) primordial mass segregation. Their differences begin to show after stellar evolution produces the first SN, the subsequent mass loss being more destructive for the initially segregated cluster.}
\label{VarSEG25k}
\end{figure*}

Fig. \ref{SEG_Ser100Ser300} shows that the impact of primordial mass segregation becomes less pronounced
with increasing cluster mass, possibly due to the evolution during the embedded ultra-compact $\ion{H}{ii}$ phase in the $0.6\ \mathrm{Myr}$ prior to gas expulsion:
the massive members would dynamically segregate (mostly due to dynamical friction) to a
greater extent with increasing cluster mass and hence
central density within this time, resulting in a mass segregated initial condition anyway
at the onset of gas expulsion.
However, the segregated clusters, where stellar evolutionary mass loss
is generally more pronounced in the inner regions, should have a smaller bound fraction
than their unsegregated counterpart.
This aspect is demonstrated in Fig. \ref{VarSEG25k} for two $2.5\times10^{4} ~ \mathrm{M}_{\odot}$ clusters;
while the evolution of their Lagrange radii is comparable right after gas expulsion,
the differences show up after $\approx3\ \mathrm{Myr}$, when the supernovae begin.
From the inner layers to outwards, the initially-segregated cluster shows an ongoing
expansion with a higher rate than the unsegregated cluster. Consequently, primordial mass segregation
would lead to a less concentrated cluster than that without it, for the same initial mass.

\begin{figure*}
\centering
\includegraphics[width=6cm, angle=-90]{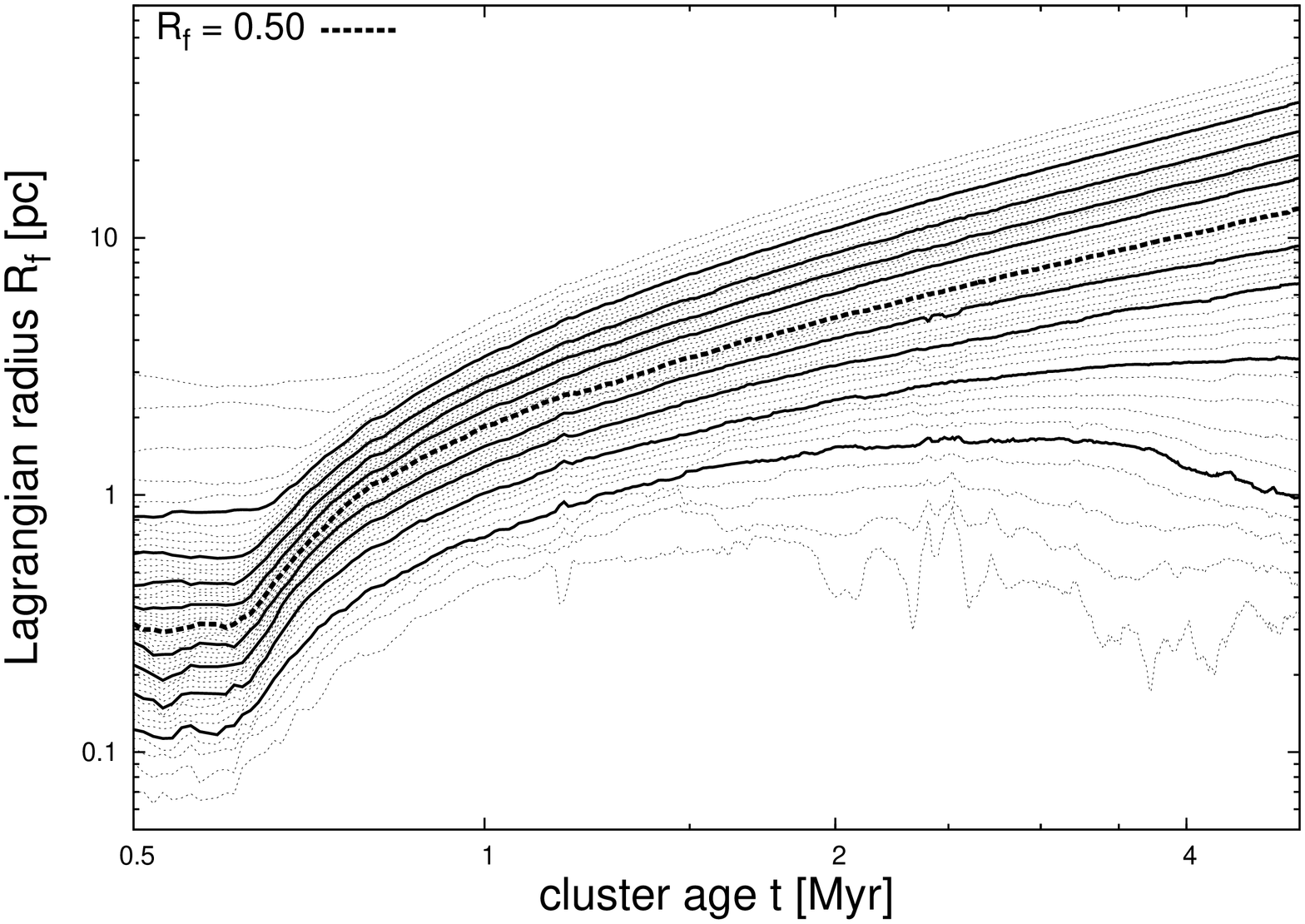}
\includegraphics[width=6cm, angle=-90]{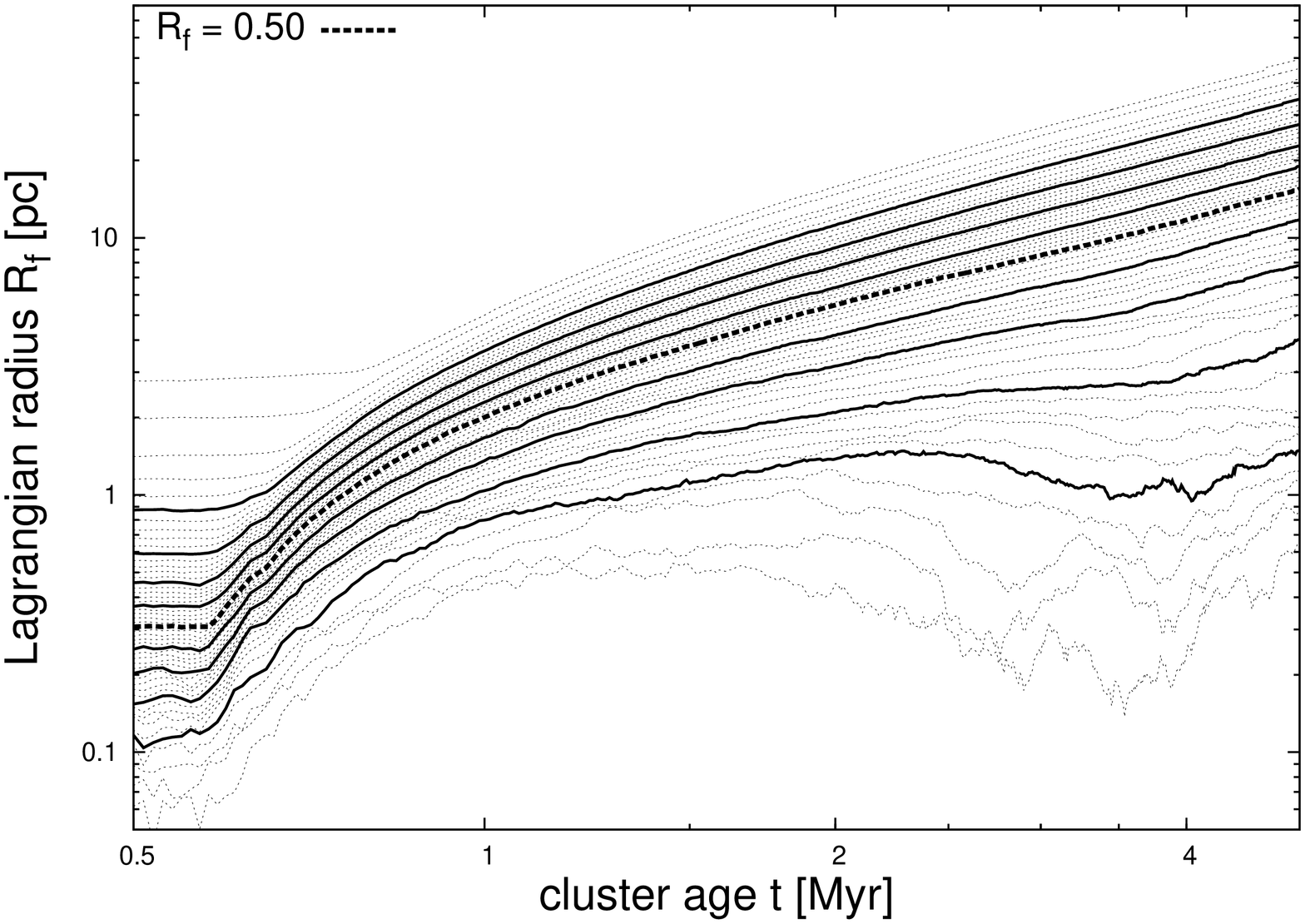}
\includegraphics[width=6cm, angle=-90]{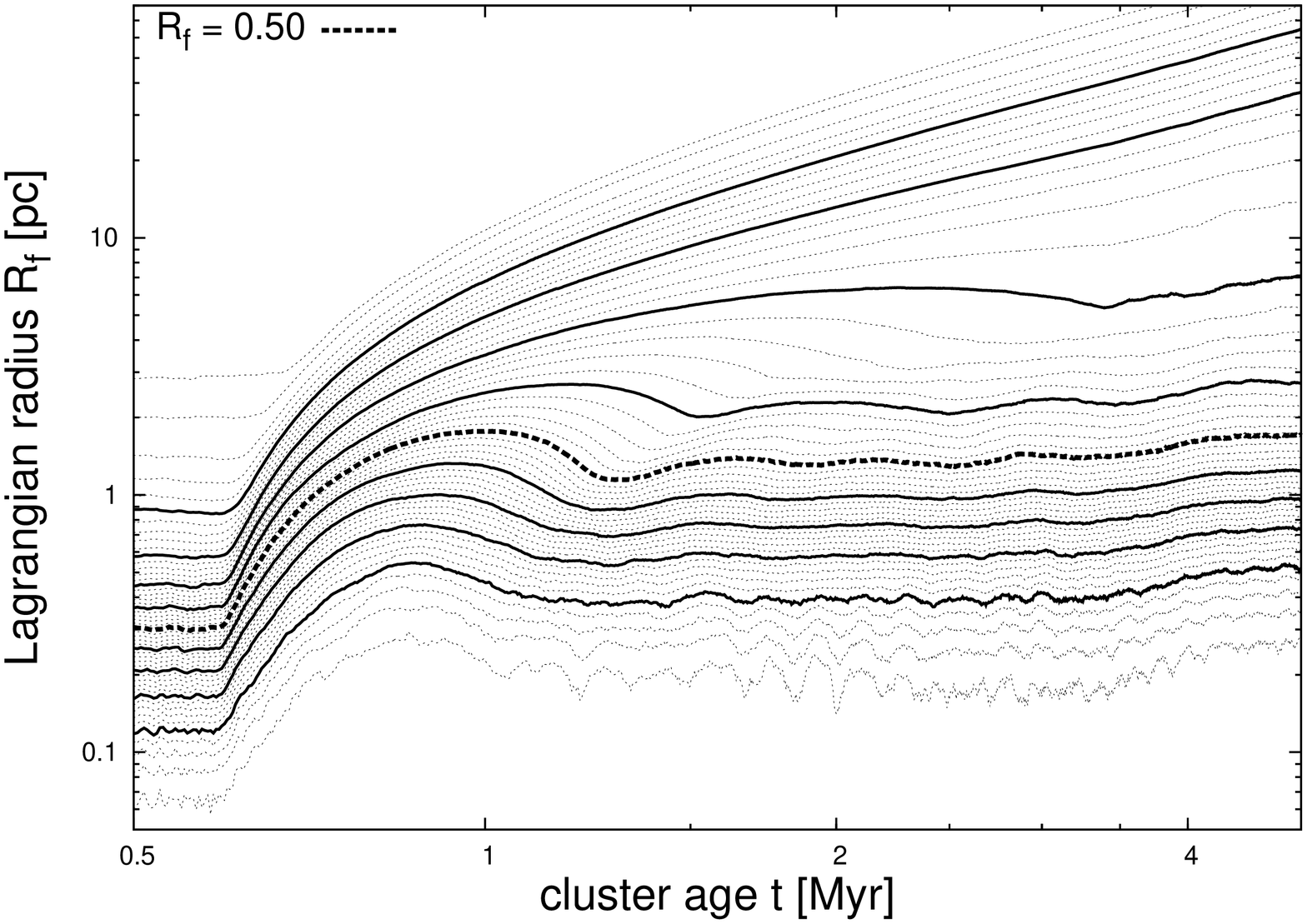}
\includegraphics[width=6cm, angle=-90]{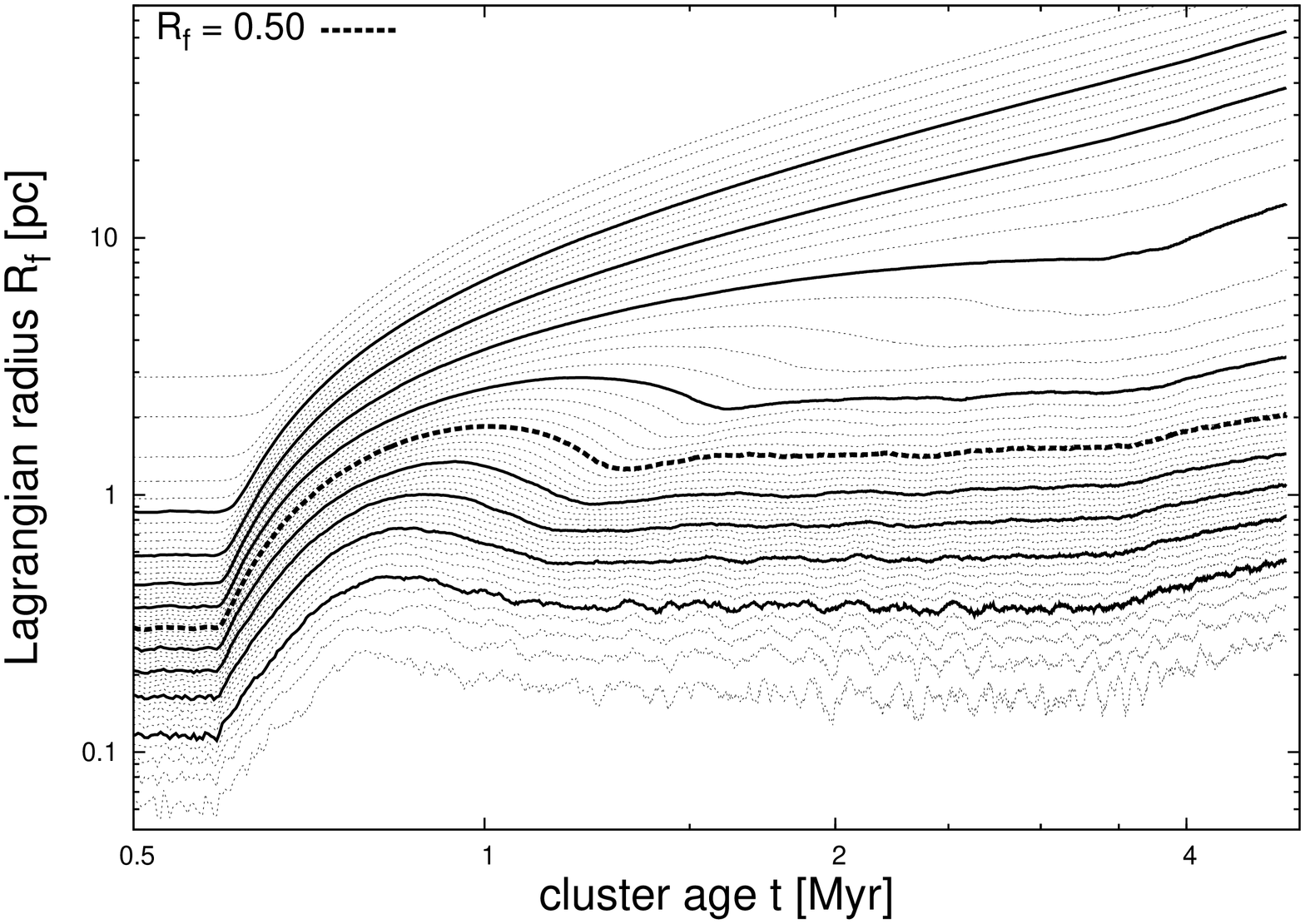}
\caption{Effect of primordial mass segregation for masses of $5\times10^{3} ~ \mathrm{M}_{\odot}$ (upper panels) and $5\times10^{4} ~ \mathrm{M}_{\odot}$ (lower panels). The plots on the left are unsegregated, the ones on the right segregated. The strong dependence of the re-virialization time on the cluster mass is obvious, but also the more destructive impact of stellar evolution on segregated clusters can be seen over the whole mass range.}
\label{fig:ex3}
\end{figure*}

The consequence of primordial mass segregation on low mass and massive clusters is considered in the simulations
illustrated in Fig. \ref{fig:ex3}. For the lower mass $5\times10^{3} ~ \mathrm{M}_{\odot}$ cluster,
the process of re-virialization is still ongoing at $t\approx4.5$ Myr, making it more vulnerable
to the additional mass loss through stellar evolution. This aspect is more profound for the segregated cluster,
where the additional stellar mass loss leads to re-expansion of the cluster.
The influence of primordial mass segregation is subtle for more massive clusters,
which largely re-virialize before the supernovae commence.
The less concentrated structure in the primordially-segregated case, as discussed above,
is still perceivable though, and it shows similar re-expansion as the lower mass cluster.

\subsection{Stellar evolution}\label{StEvo}

It is intuitively clear that mass loss through stellar evolution will cause the cluster potential
to dilute, aiding the loss of stars from the cluster. Although including stellar evolution is
physically realistic, we do additional simulations without it, so that we can assess its impact.
These simulations cover initially mass-segregated clusters with masses between
$5\times10^{3} ~ \mathrm{M}_{\odot}$ and $5\times10^{4} ~ \mathrm{M}_{\odot}$.

\begin{figure*}
\centering
\includegraphics[width=6cm, angle=-90]{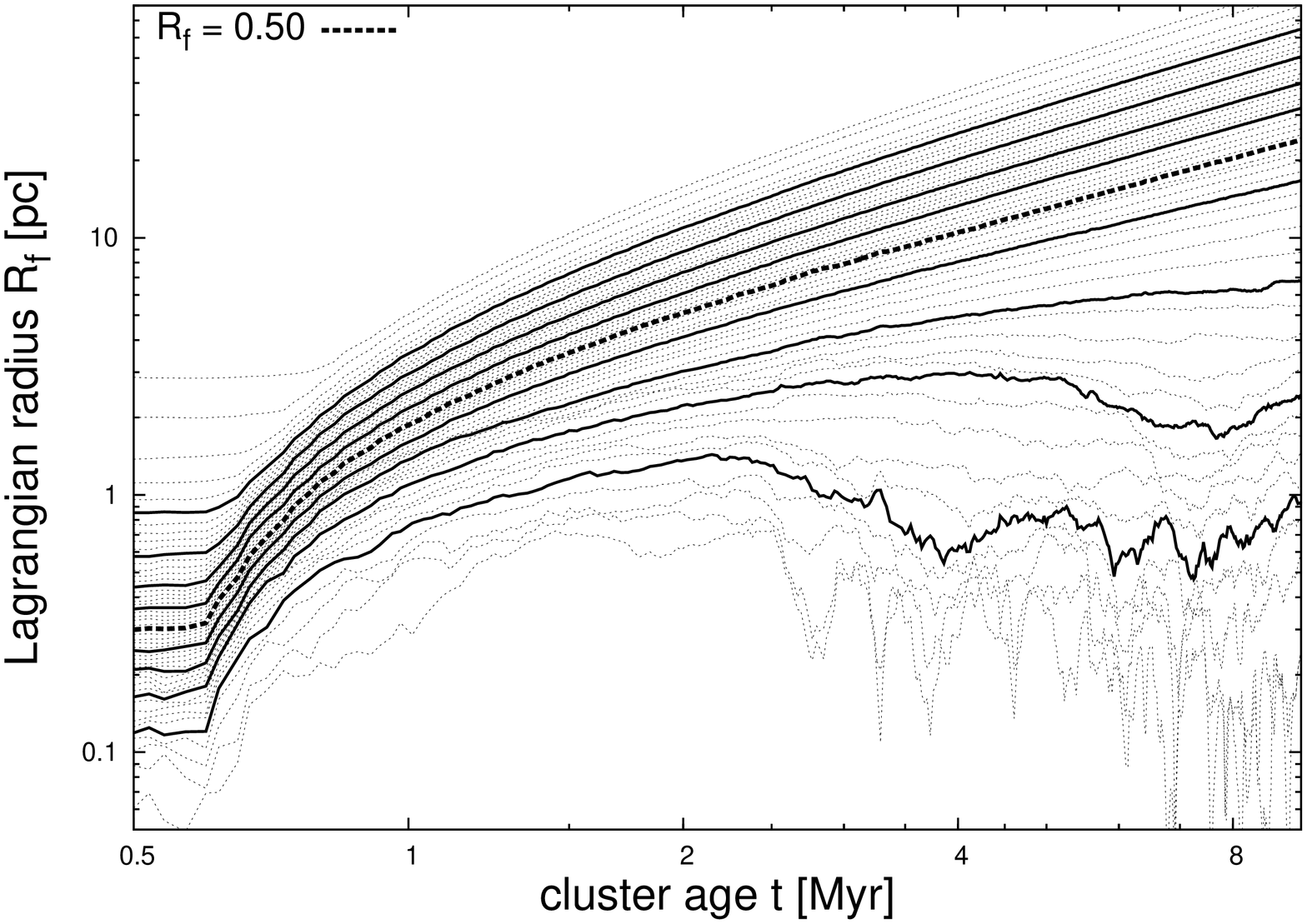}
\includegraphics[width=6cm, angle=-90]{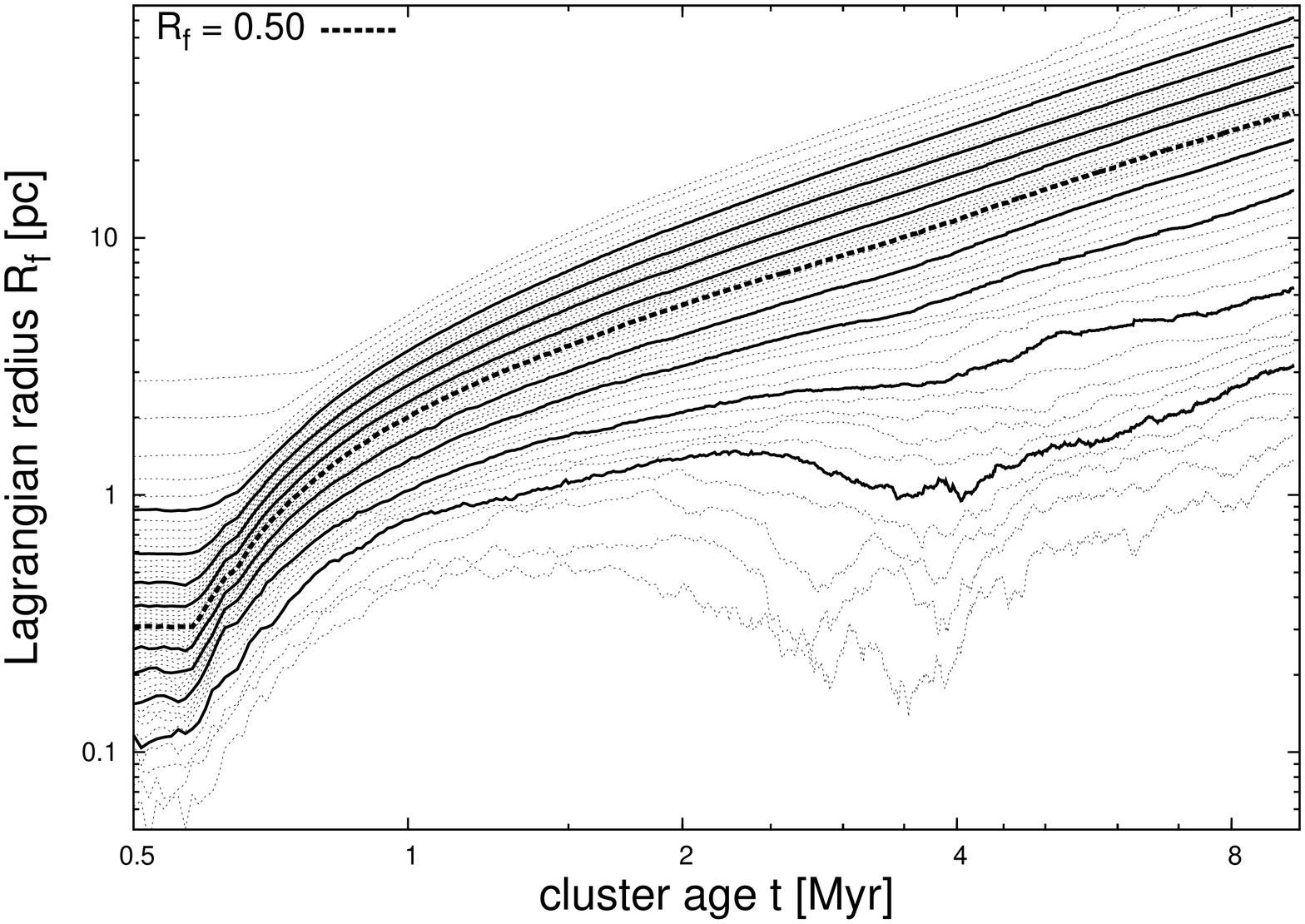}
\caption{Effect of stellar evolution. Covering the first $9.5\ \mathrm{Myr}$ (respectively $8.9\ \mathrm{Myr}$ after the beginning of gas expulsion) for two $5\times10^{3} ~ \mathrm{M}_{\odot}$ clusters. While the reversion process after gas expulsion is ongoing for the cluster without stellar evolution (left panel), the first SN disturb the process of re-virialization in the more realistic scenario including stellar evolution (right panel). The effect is more distinct in these lower mass clusters.}
\label{StEvo5k}
\end{figure*}

The difference caused by stellar evolution is demonstrated in Fig. \ref{StEvo5k} for lower mass clusters,
most notably when the stellar mass loss boosts after the first SN at $\approx 4$ Myr. After $9.5\ \mathrm{Myr}$,
even the innermost layers of the model cluster including stellar evolution are still expanding considerably,
having only $\approx2$\% of the initial cluster mass left within a radius of $\approx 1 ~ \mathrm{pc}$.
All layers, including those which initially rebound after gas expulsion, show an ongoing expansion.
In contrast, the inner layers of the model without stellar evolution are still within that radius
and do not expand significantly.

\begin{figure}
\resizebox{\hsize}{!}{\includegraphics[angle=-90]{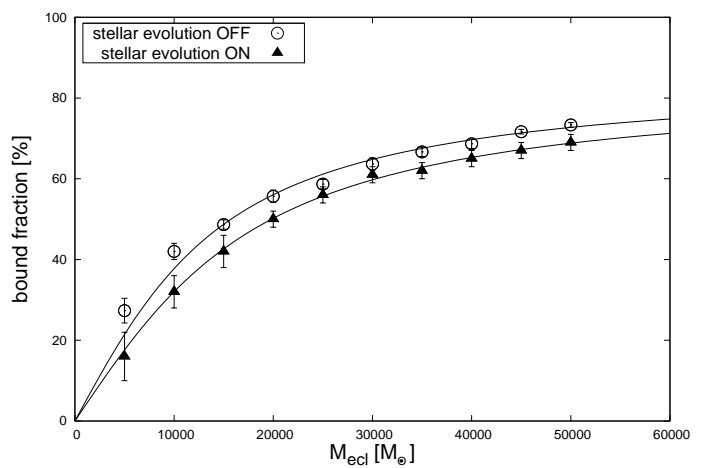}}
\caption{The influence of stellar evolution on the bound fraction as a function of initial cluster mass. The effect lessens with increasing mass, but is noticeable over the whole examined mass range. The differences are most likely due to the different re-virialization times after gas expulsion. The values for the clusters without stellar evolution are averaged over 3 simulations per mass, while one simulation per mass is used for those including stellar evolution.}
\label{StEvo1}
\end{figure}

The overall influence of stellar evolution is summarized in fig. \ref{StEvo1},
it illustrates that stellar evolution reduces
the bound fraction over the whole examined mass range. The impact on the lower mass clusters seems
to be larger, due to their longer re-virialization time and
thereby higher sensitivity to the additional stellar mass loss.
Consequently, not only the bound fraction, but the structure of the inner parts of the
surviving cluster can be strongly influenced by stellar evolution.
For the regarded HMR $0.3 ~ \mathrm{pc}$ clusters, stellar evolution reduces the bound fraction
for about 10\% in the lower mass regime, and stagnates just under 5\% for higher masses.

\subsection{Stellar evolution vs. dynamical friction}\label{dyfric}

It would be worthwhile to consider how the bound fraction is affected by the dynamical
heating effect alone, arising from the dynamical friction of the most massive stars.
Without the stellar evolution and hence the additional potential drop through
stellar mass loss,
the bound fraction would tend to increase. However, without the mass loss
(in other words, if the stars were point masses)
the most massive stars, which are $\gtrsim100\Ms$, would as well deposit energy preferentially
to the central part of the cluster, due to their repeated scattering from and sinking back
to the core via dynamical friction. The importance of the above effect
can be assessed in models that maximize it, \ie, in initially mass-segregated clusters
without stellar evolution.

\begin{figure}
\resizebox{\hsize}{!}{\includegraphics[angle=-90]{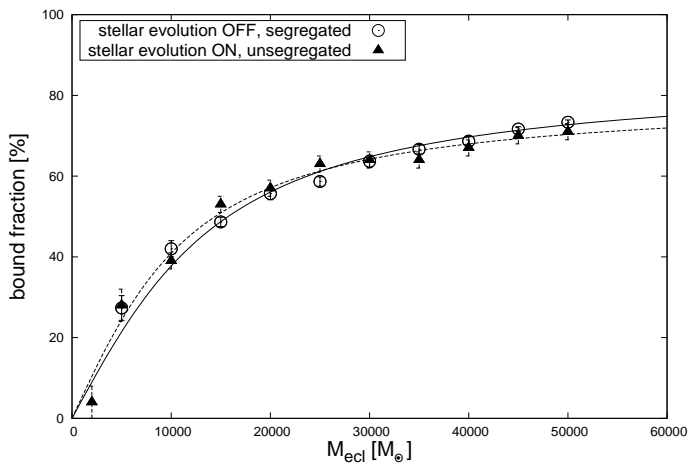}}
\caption{The impact of dynamical friction, estimated through minimization and maximization of its effect. The overlap over large parts of the mass range suggests that its heating of the cluster might be of the same order of magnitude as the influence of stellar evolution. The simulations maximizing dynamical friction are averaged over 3 runs per mass, the data points for minimized dynamical friction correspond to one realization per mass.}
\label{PlotDynamicalFriction}
\end{figure}

This is summarized in Fig. \ref{PlotDynamicalFriction} which shows that the
maximal effect of heating of the cluster through dynamical friction alone
could be of
the same order as that due to stellar evolution has on
the bound fraction for an unsegregated cluster.

\begin{figure*}
\centering
\includegraphics[width=6cm, angle=-90]{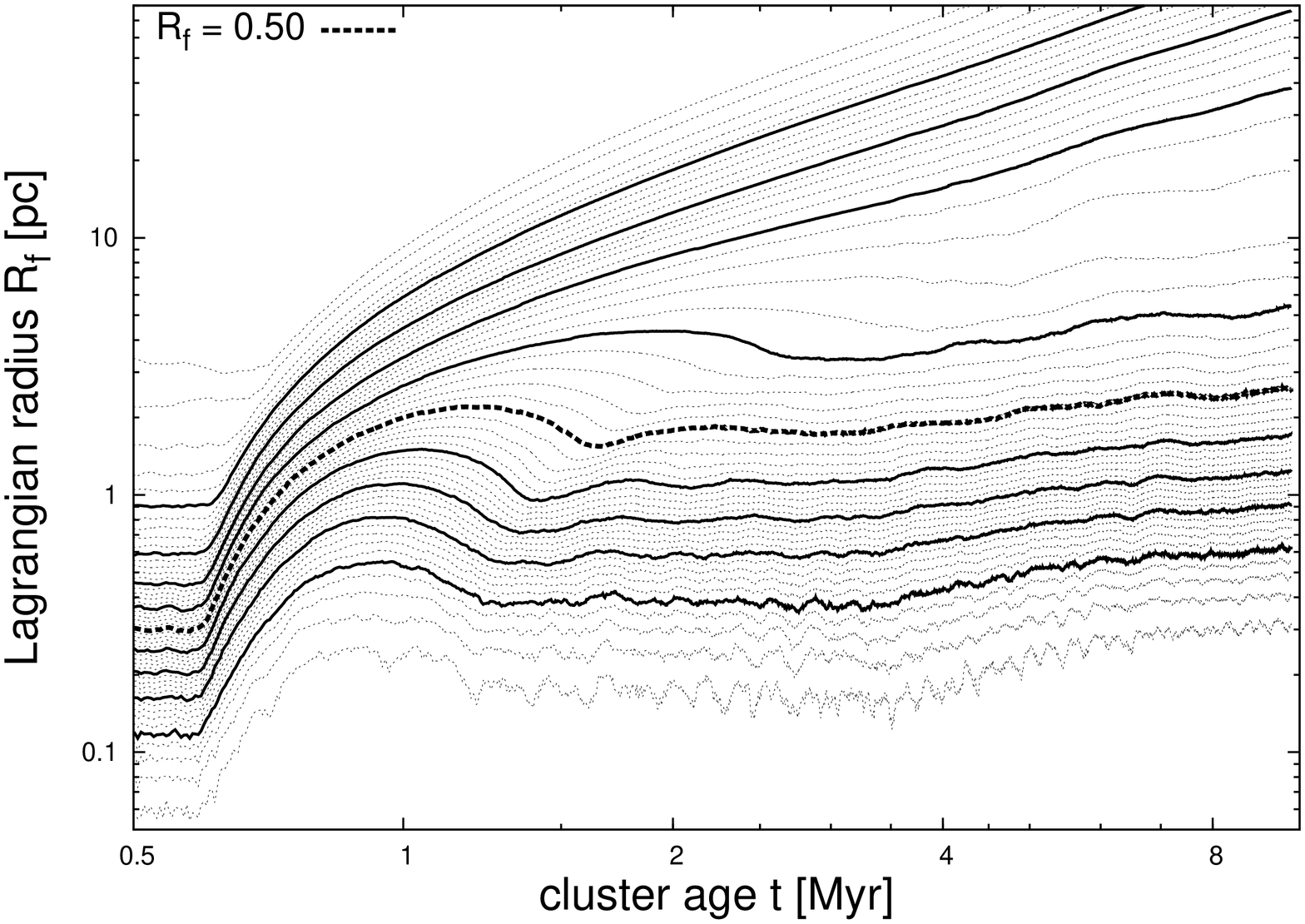}
\includegraphics[width=6cm, angle=-90]{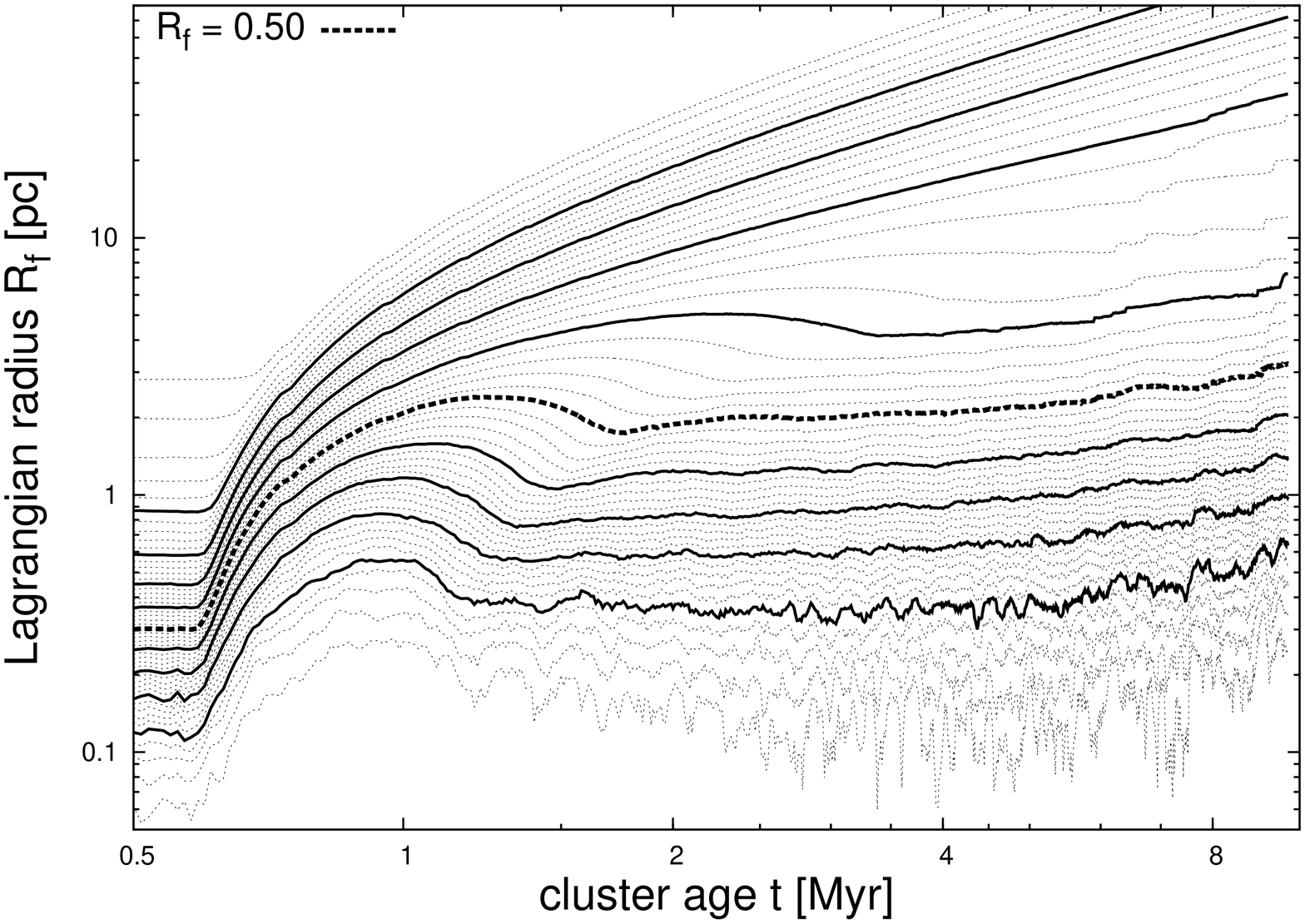}
\caption{Impact of dynamical friction, gauged with exemplary plots, tracking the first $10\ \mathrm{Myr}$ (respectively $9.4\ \mathrm{Myr}$ after the beginning of gas expulsion) for two \(3\times10^{4}\) \(\mathrm{M}_{\odot}\) clusters. The bound fraction and overall extension are of the same order of magnitude at that time. The expansion of the cluster with minimized dynamical friction (left panel) is driven primarily by stellar evolution, whereas the interactions through dynamical friction provide a constant source of heating for the other cluster (right panel).}
\label{DynFricComp}
\end{figure*}

A comparison between two $3\times10^{4} ~ \mathrm{M}_{\odot}$
clusters is shown in Fig. \ref{DynFricComp}, illustrating that neither
the overall expansion nor the bound fraction is notably different.
The expansion due to stellar mass loss via supernovae in one cluster (left panel),
starting at $\approx3.5\ \mathrm{Myr}$,
and the flattening at about $7\ \mathrm{Myr}$ when the cluster
is close to dynamical equilibrium again,
are of the same order as the heating through dynamical friction alone.
The innermost parts of the model cluster with maximized dynamical friction (right panel)
show stronger fluctuations, the interactions with the still existing most massive
particles providing a persistent source of heating.

\subsection{The tidal field at solar distance}\label{tfsolar}

The clusters simulated so far were assumed to be isolated, \ie, without the influence
of an external galactic tidal field. Their corresponding bound fractions therefore represent an
upper limit to that for the more realistic scenario that includes the external forces.

Here we consider the tidal field at the solar distance, approximated by the field of a
point mass of $2\times10^{10} ~ \mathrm{M}_{\odot}$ at a distance of
$8.5 ~ \mathrm{kpc}$, orbited by the model clusters with a circular
velocity of $220 ~ \mathrm{km~s^{-1}}$.

\begin{figure}
\resizebox{\hsize}{!}{\includegraphics[angle=-90]{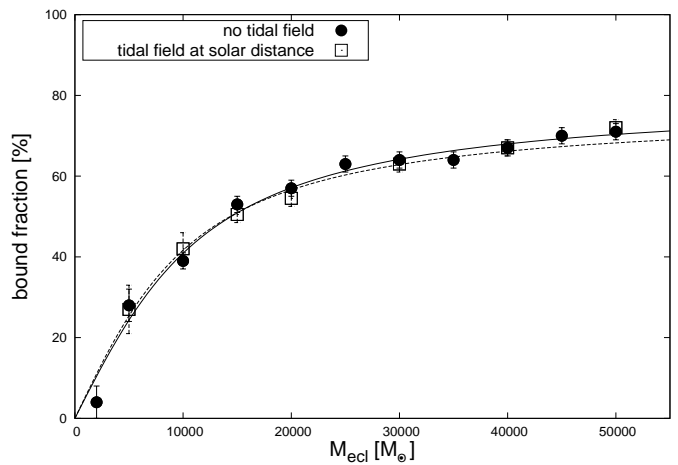}}
\caption{The impact of the tidal field at solar distance on the bound fraction as a function of initial cluster mass. With their relatively compact initial HMR of 0.3 pc, the clusters show no distinct reaction to the field. Every data point corresponds to one simulation for the isolated clusters, the data points for the clusters in a tidal field consider 2 simulations per mass.}
\label{TidalFieldSolar}
\end{figure}

The results are shown in Fig. \ref{TidalFieldSolar} which suggest that
the clusters remain well within their tidal radii at all times, for solar-like Galactocentric
distance, making the influence of the tidal field negligible
for the range of cluster mass and HMR and the $\varepsilon_\mathrm{SFE}$ considered here.

\subsubsection{Removing escaped stars}

\begin{figure}
\resizebox{\hsize}{!}{\includegraphics[angle=-90]{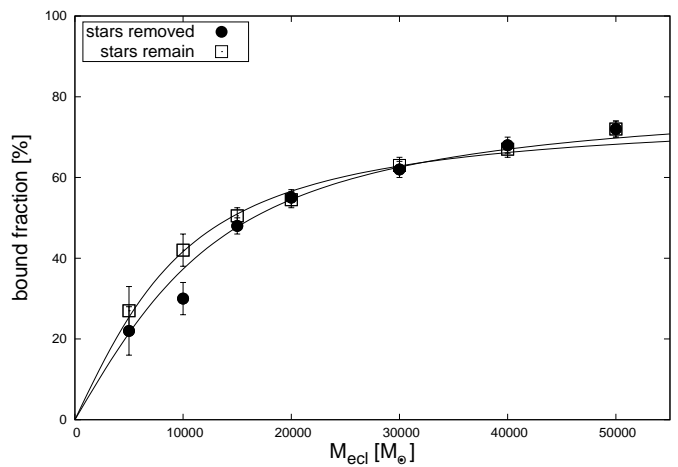}}
\caption{Comparison between removing and keeping escaped stars in the calculations. Removing escapers from clusters situated in a tidal field at the solar distance does not seem to have a significant impact on the bound fraction over the considered mass range. The overlap with the comparison plot including the escapers in the total cluster mass is notable in wide parts. For the clusters with removed stars, one simulation per mass is used, while the data points for the clusters keeping escapers consider 2 runs per mass.}
\label{TidalRemSolar}
\end{figure}

Unlike the question concerning the importance of incorporating stellar
evolution discussed in \ref{StEvo}, the removal of stars which are no
longer bound to the cluster from the calculations is of a more technical interest.
If the bound fraction after gas expulsion does not change either way,
it is more economical to remove them from the calculations. The results for a tidal field at solar distance
shown in Fig. \ref{TidalRemSolar} indicate that the removal of escapers is indeed viable,
assuming our examined mass range and HMR.

\subsection{Strong tidal field: A simple model of the Arches Cluster}\label{Arches}

While a Galactic tidal field does not seem to have a significant impact on
the bound fraction at the solar distance, clusters much closer to the
Galactic center could be influenced due to the much stronger tidal field there and hence smaller
tidal radii. In order to estimate the effect of a strong tidal field,
we compute a simplified model of the Arches cluster and try to reconstruct its initial mass and HMR.

The cluster is set to orbit the Galactic center at a distance of
$100 ~ \mathrm{pc}$ (an estimate between the $30 ~ \mathrm{pc}$ projected
distance from \citealt{Nagata1995} and the upper limit of
$200 ~ \mathrm{pc}$ suggested by \citealt{Stolte2008}) with an orbital velocity of $200 ~ \mathrm{km~s^{-1}}$,
comparable to the values mentioned in \citet{Stolte2008} or \citet{Clarkson2012}.
It encloses a point mass of \(10^{9}\) \(\mathrm{M}_{\odot}\) (a rough estimate, accounting for the
central SMBH and parts of the nuclear bulge with a total mass of
$\approx \left( 1.4 \pm 0.6 \right) \times10^{9} ~ \mathrm{M}_{\odot}$ \citep{Launhardt2002}).
The previous simulations here show an expansion of the inner Lagrange radii by a factor of about
3 to 4 after re-virialization for high-mass, compact clusters, in quite good agreement with the theoretical estimate
\citep{Kroupa2008initial} for adiabatic gas removal 
\begin{equation*}
 \frac{r_\mathrm{after}}{r_\mathrm{initial}} = \frac{1}{\varepsilon_\mathrm{SFE}}  = 3
\end{equation*}
for $\varepsilon_\mathrm{SFE}$ 0.33.
So from the measured HMR of around $0.4 ~ \mathrm{pc}$ (\citealt{Olczak2012}
and references therein) one can estimate the approximate initial HMR being
$0.1 ~ \mathrm{pc}$ - $0.15 ~ \mathrm{pc}$.
We vary the initial mass and compare the resulting HMR after 2.5 Myr,
which corresponds to the current age of the Arches Cluster \citep{Figer2002}.

\begin{figure*}
\centering
\includegraphics[width=6cm, angle=-90]{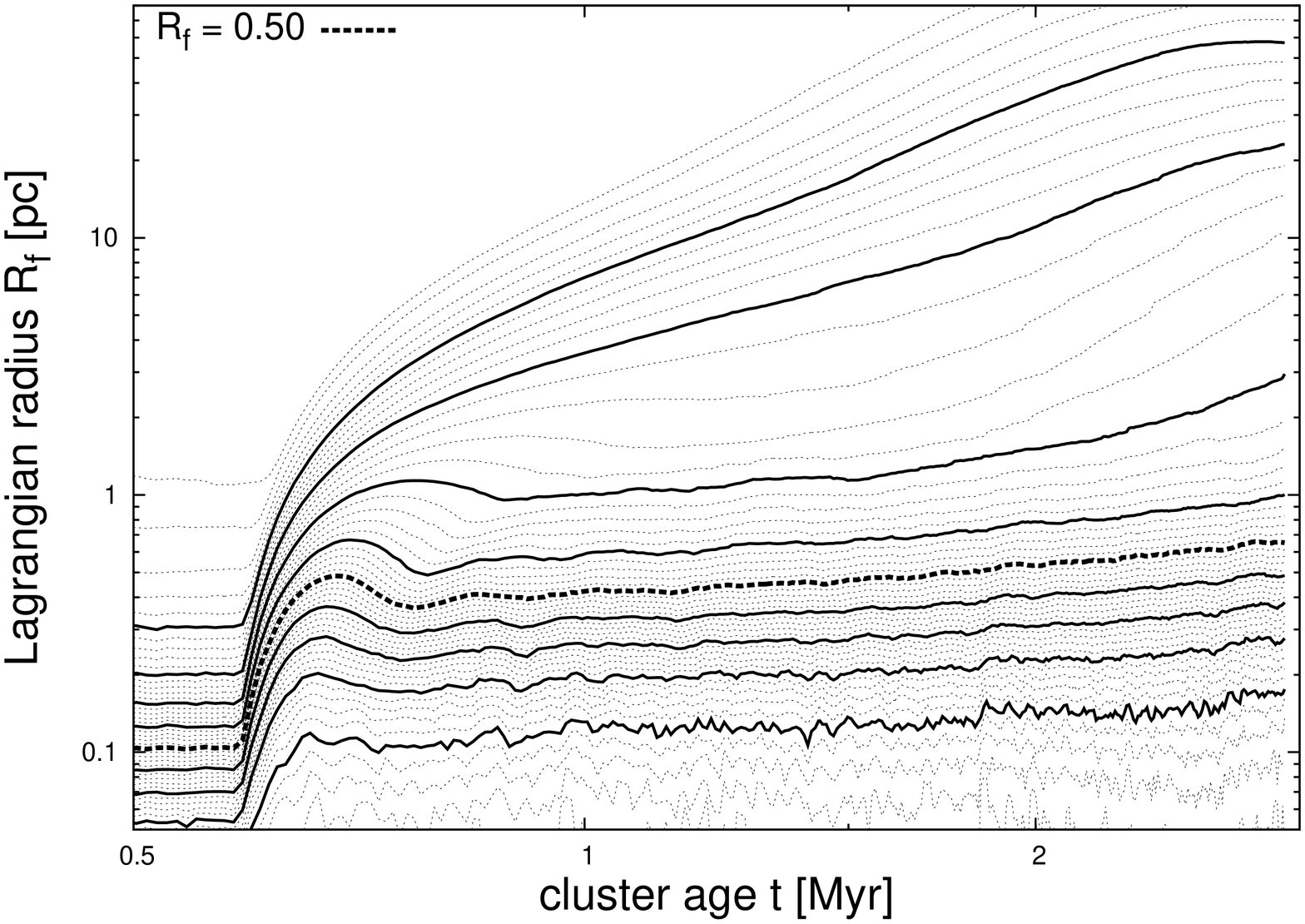}
\includegraphics[width=6cm, angle=-90]{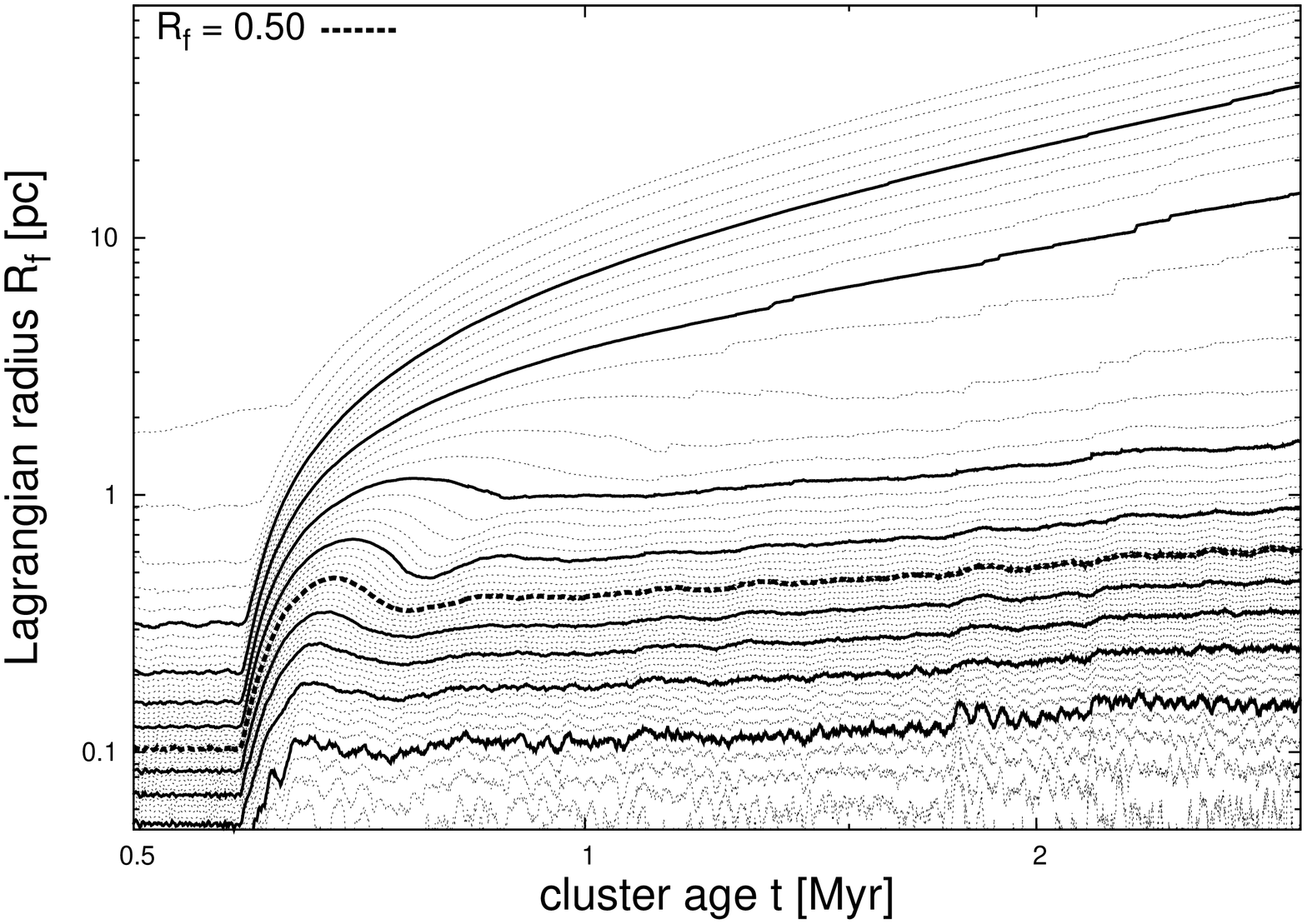}
\caption[Tidal field near the GC: Arches cluster]{Comparing two $3\times10^{4} ~ \mathrm{M}_{\odot}$ clusters, both with an initial HMR of $0.1 ~ \mathrm{pc}$. The left panel shows a simplified model of Arches in a strong Galactic tidal field, the right model is considered to be in isolation.}
\label{ArchesCompare}
\end{figure*}

\begin{figure}
\resizebox{\hsize}{!}{\includegraphics[angle=-90]{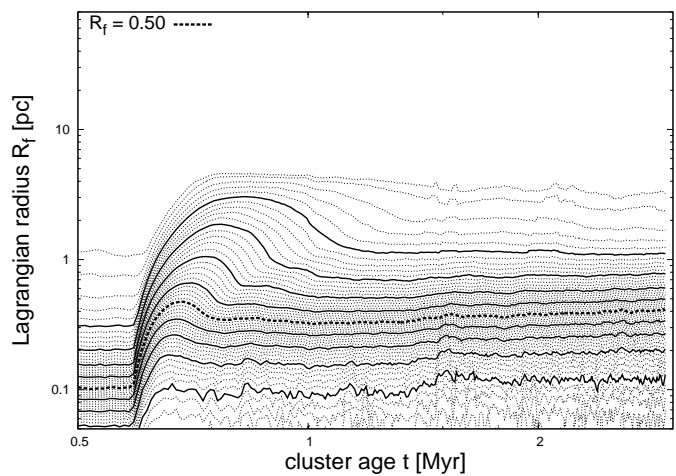}}
	\caption{The tidal radius in the strong tidal field is emphasized by the removal of escaped stars from the calculations. The plot shows a $3\times10^{4} ~ \mathrm{M}_{\odot}$ cluster with an initial HMR of $0.1 ~ \mathrm{pc}$.}
	\label{ArchesRem}
\end{figure}

As to be expected, the computed clusters show a reaction to the strong tidal field (see Fig. \ref{ArchesCompare}), the slope of their outer Langrange radii being steeper, their overall expansion after gas expulsion bigger. The inner structure of both clusters looks similar though, suggesting that the influence of the external field can be regarded as being limited to the outer layers in good approximation. Outer parts of the cluster which initially reverse within 1 Myr experience a strong outward acceleration once their distance to the cluster center exceeds $\approx$ 3 to $4 ~ \mathrm{pc}$ (representing the tidal radius of the cluster, illustrated in Fig. \ref{ArchesRem}). It is noteworthy though that the bound fraction does not differ over more than $\approx$ 10\% with or without the tidal field, which is most likely due to the very compact initial size of the clusters. However, the ongoing albeit moderate overall expansion after gas expulsion results in a HMR over $0.4 ~ \mathrm{pc}$ after just 2.5 Myr, even for the heaviest simulated clusters of $3.5\times10^{4} ~ \mathrm{M}_{\odot}$. This means that one would need a more massive, but preferably an even more compact cluster with an initial HMR smaller than 0.1 pc to match the parameters of the Arches cluster given further below.

\begin{figure}
\resizebox{\hsize}{!}{\includegraphics[angle=-90]{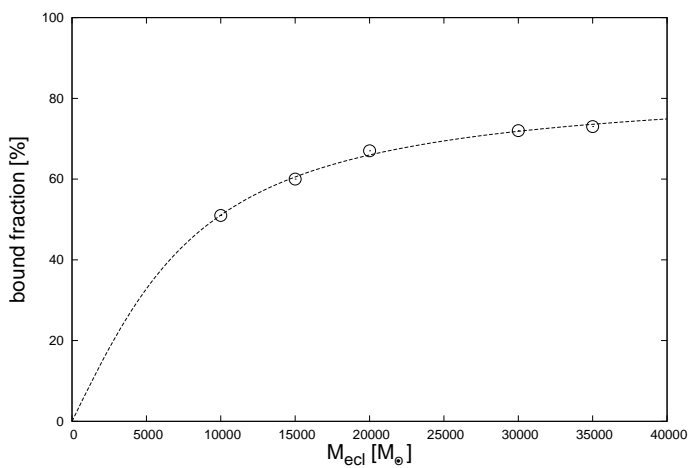}}
\caption{The mass fraction within 4 pc as an estimate of the bound fraction for the Arches Cluster at its current age of 2.5 Myr and an initial HMR of $0.1 ~ \mathrm{pc}$.}
	\label{Arches4pc}
\end{figure}

\begin{figure}
\resizebox{\hsize}{!}{\includegraphics[angle=-90]{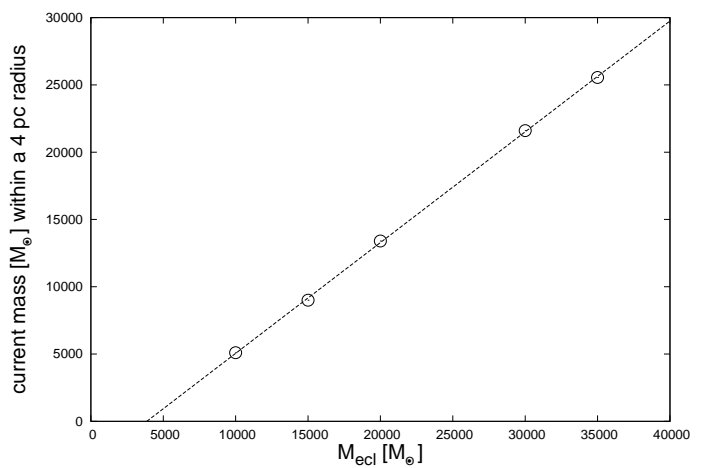}}
	\caption{The mass within 4 pc for more convenient comparisons with observational data of the Arches cluster. Measurements of the current mass point to the initial mass $M_\mathrm{ecl}$, provided the other parameters match the actual conditions close enough.}
	\label{ArchesCurrentMass}
\end{figure}

As an approximation of the bound fraction, we now look at the Lagrange radii after 2.5 Myr and determine what layers are within $4 ~ \mathrm{pc}$ (a rough estimate of the tidal radius, which will slightly overestimate the bound fraction). So in this case, we will not determine the maximum bound fraction, but the mass fraction within a certain radius at a given time. The results are shown in Fig. \ref{Arches4pc}. To make comparisons with observational data easier, Fig. \ref{ArchesCurrentMass} contains the same information as Fig. \ref{Arches4pc}, but shows the current mass instead of the bound fraction. The current mass measured within 3$\arcmin$ (respective extent for an observer from Earth) would then point directly to the initial cluster mass, should the other parameters (initial HMR, $\varepsilon_\mathrm{SFE}$, etc.) be realistic.

Estimates for the current mass and possible initial mass vary for different studies. A current mass of $ \left( 3.1 \pm 0.6 \right) \times10^{4} ~ \mathrm{M}_{\odot}$ \citep{Espinoza2009} suggests an initial mass of just over $4 \times10^{4} ~ \mathrm{M}_{\odot}$ from our simulations, whereas a current mass of $ \left( 1.9 \pm 0.3 \right) \times10^{4} ~ \mathrm{M}_{\odot}$ suggested by \citet{Habibi2013} prompts an initial mass of around $2.7 \times10^{4} ~ \mathrm{M}_{\odot}$. The aforementioned underestimated growth of the HMR in 2.5 Myr would still warrant the assumption of a smaller initial HMR, but the already carried out simulations portray a reasonable approximation. Estimates of the initial mass vary from around $2\times10^{4} ~ \mathrm{M}_{\odot}$ \citep{Kim2000}, over $\left( 4.9 \pm 0.8 \right) \times10^{4} ~ \mathrm{M}_{\odot}$ \citep {Harfst2010} to the upper mass limit of $\sim 7\times10^{4} ~ \mathrm{M}_{\odot}$ (within $0.23 ~ \mathrm{pc}$) of \citet{Figer2002}.

Our result seems to agree quite well with the above values, although we make several approximations. The actual environment around Arches is much more complex, but nevertheless the incorporation of a strong tidal field shows that an Arches-like cluster can readily survive gas expulsion even under these conditions.

\subsection{Radial velocities}
\begin{figure*}
\centering
\includegraphics[width=6cm, angle=-90]{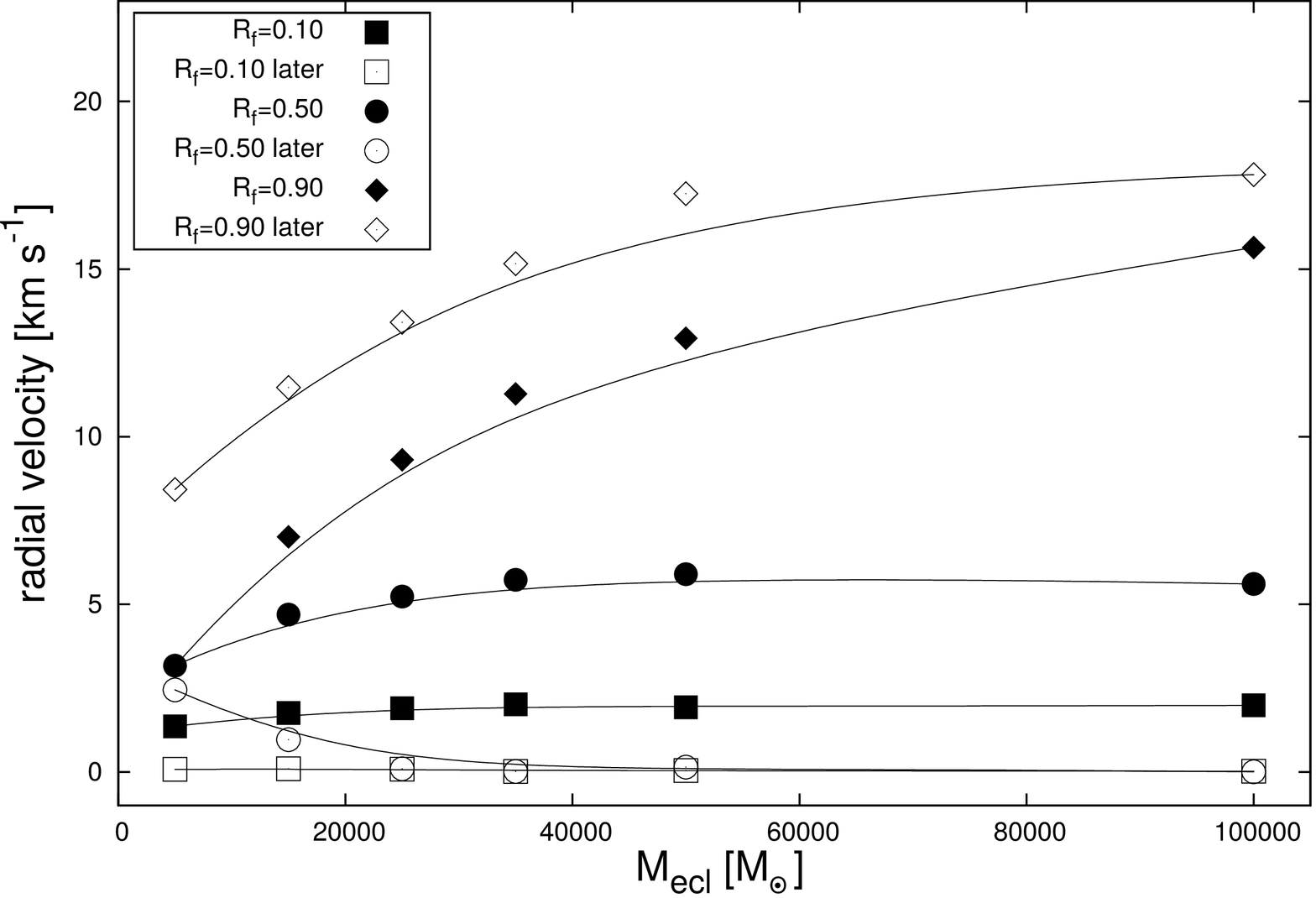}
\includegraphics[width=6cm, angle=-90]{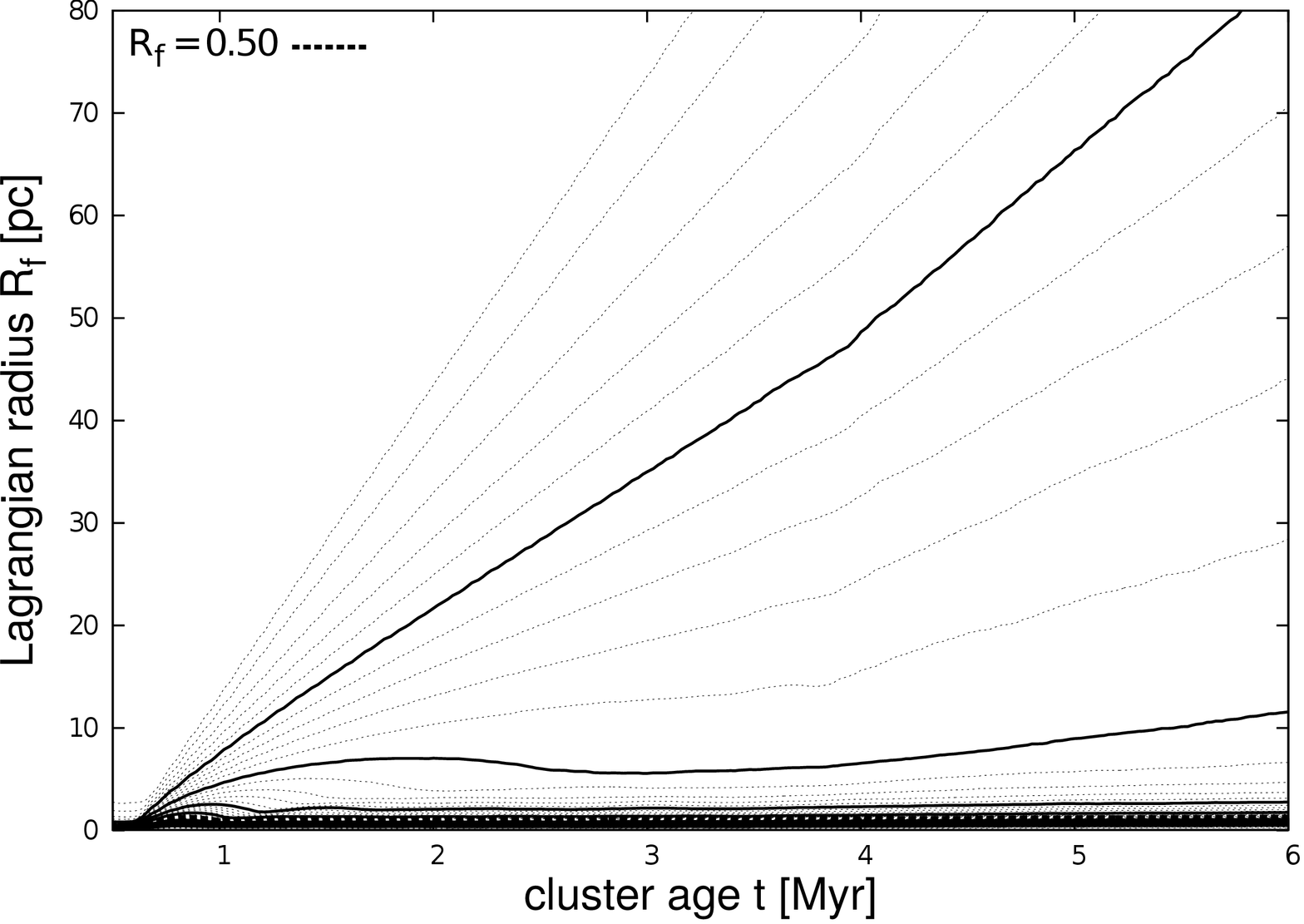}
\caption{The left panel shows the radial velocities right after gas expulsion and at a cluster age around $6\ \mathrm{Myr}$ as a function of $M_\mathrm{ecl}$. The simulations are for clusters with $\varepsilon_\mathrm{SFE}=0.33$ and initial HMR of $0.3 ~ \mathrm{pc}$. They consider stellar evolution, no primordial mass segregation and no Galactic tidal field, one simulation per mass. The right panel shows the evolution of Lagrange radii for a  $10^{5} ~ \mathrm{M}_{\odot}$ cluster (this time not in logarithmic scale, resulting in a clear visibility of only the outermost radii) to explain the acceleration of the outer layers: at a cluster age around $3.8\ \mathrm{Myr}$, stellar evolution causes noticeable mass loss. The outer low density regions have not yet slowed down from the initial expansion due to gas expulsion, because the stars have little chance to interact. Further mass loss through stellar evolution can therefore not be compensated, resulting in an accelerated expansion.}
\label{VelEx}
\end{figure*}

Apart from determining the bound fraction after gas expulsion, our simulations enable a rough estimate for the radial velocities of the unbound outer layers. Determining those velocities might help to understand the heating of the Galactic thick disc (\citep{PavelDisc,AssmannFellhauer2011}). If one can find a relation between the radial velocity and the initial mass, it may be possible to constrain values for cluster masses responsible for properties of the Galactic thick disc. The observed chain galaxies at high redshift \citep{Elme2006} suggest that this model of star-cluster-birth induced heating of galactic discs may indeed be relevant.\\
Fig. \ref{VelEx} gives an impression of the radial velocities of inner and outer layers, right after gas expulsion and at a cluster age around $6\ \mathrm{Myr}$. The velocities right after gas expulsion are averaged over a cluster age of $0.6\ \mathrm{Myr}$ to $0.7\ \mathrm{Myr}$, the later velocities are averaged over $6\ \mathrm{Myr}$ to $6.5\ \mathrm{Myr}$. The latter point in time is somewhat arbitrary and chosen because the re-virialization process is completed for most clusters, yet the low mass clusters are not yet totally disrupted by stellar evolution.

One can see that the expansion of the inner layers (10\% and 50\%) has slowed down or even stopped after $6\ \mathrm{Myr}$, whereas the outer 90\% layer accelerated further. The latter may be caused by stellar evolution: the few two-body encounters in the outer parts with low density can not slow down the expansion effectively, so that the additional mass loss through stellar evolution causes an outward acceleration (see Fig. \ref{VelEx}). A viable estimate of a sufficiently high cluster mass to explain properties of the thick disc will need more simulations and should be addressed in future studies. \citet{PavelDisc} suggests  a cluster mass of $10^{6} ~ \mathrm{M}_{\odot}$, which is currently too massive for direct N-body calculations. The required radial velocity (or velocity dispersion perpendicular to the Galactic disc plane) of around $40 ~ \mathrm{km~s^{-1}}$ is not reached by any of our simulated clusters, but lies well beyond $10^{5} ~ \mathrm{M}_{\odot}$. With the importance of $\varepsilon_\mathrm{SFE}$ implied by Fig. \ref{CompSFE}, the required mass could be reduced for clusters with lower star-formation efficiency. Additionally, external fields could accelerate the outer layers yet more, thus also implying a time dependence of the radial velocities even at later cluster age.

\section{Discussion and conclusions}
Several commonly used approximations are made in our simulations. For instance, the system is in equilibrium at the start of gas expulsion, we do not consider sub-structures, there are no primordial binaries and $\varepsilon_\mathrm{SFE}$ is constant over the whole cluster.\\
The self-consistent formation of a bound cluster from a giant molecular cloud carried out by \citet{HurleyBekki2008} shows that equilibrium is reached in the order of a few crossing times, with said timescale being fairly short for our compact configurations and well within the adopted $\tau_\mathrm{d}\approx 0.6\ \mathrm{Myr}$.

Furthermore, this simplifying assumption is necessary to make the models computable, because simulating a forming cluster of mass larger than about $100 ~ \mathrm{M}_{\odot}$ magneto-hydrodynamically with feedback is beyond the current technological capabilities. In reality the earliest phases of embedded cluster assembly are given by the simultaneous inward flow of gas, driven by the deepening potential well of the embryonic embedded cluster, and by the nearly-simultaneous outflow of gas driven by the increasing stellar feedback and magnetic fields generated in the proto-stellar accretion disks (a single proto-star taking $\approx 10^{5} ~ \mathrm{Myr}$ to assemble $\approx 95 ~ \%$ of its mass, see \citealt{WuchterlTscharnuter2003}). At any particular time the existing proto-stellar population virialises within several crossing times which is very short in the embryonic embedded cluster such that the approximation of a virialised stellar population embedded in gas is more accurate than assuming the entire final population is in a free fall as if the stars formed instantaneously with little motions.
For example \citet{Dale2015} perform a set of hydrodynamical plus feedback simulations but without magnetic fields to test the general evolution of clusters of different mass. They conclude that feedback has little disruptive effect on the forming clusters.
However, in contrast to their statements, the resolution achieved is not adequate to address the issue of cluster survival or disruption adequately, because for example at the high-mass end (their initial clouds near $10^{6}~ \mathrm{M}_{\odot} $), their stellar particles have masses of $100 ~ \mathrm{M}_{\odot}$ thereby constituting unresolved substantial sub-clusters. The least massive clouds simulated ($10^{4}~ \mathrm{M}_{\odot}$) have a resolution of $1 ~ \mathrm{M}_{\odot}$ such that the embedded clusters form from unresolved sub-clusters of $1~ \mathrm{M}_{\odot}$ each.  Especially at the low-mass end, where the dynamics is heavily relaxation driven, very high precission integration techniques are required to account for the interparticle forces. \citet{Dale2015} write that in their simulations locally, where stellar particles are formed, the star-formation efficiency is near 100\%. This contradicts our assumption that $\varepsilon_\mathrm{SFE}$ is typically  $0.33$ throughout the embedded cluster volume. However, it is also in contradiction with the results of the high-resolution magnetohydrodyamical calculations by \citet{MM2012} (see also \citealt{Bate2014}) of forming individual proto-stars. \citet{MM2012} find that the magnetic fields induced in the evolving circum-stellar disk drives a bi-polar outflow such that the star-formation efficiency is about $0.26$ - $0.54$ only. Therefore, the Nbody models of embedded clusters used here capture the important physical processes.
Despite the shortcomings of either approach, the models presented here agree broadly with those of \citet{Dale2015} (their Fig. 9) since for embedded cluster masses larger than $10^{4} ~ \mathrm{M}_{\odot}$ the fraction of stars which become unbound to the removal of residual gas becomes $0.30$ or less for star formation-efficiencies of $0.30$ or more. The reason for this high fraction of bound stars despite the low star-formation efficiency, even in the present Nbody models, is that the velocity of gas expulsion is comparable to the pre-gas expulsion velocity dispersion in our models.
The overall conclusion, in broad but not in detail agreement with those of \citet{Dale2015}, is thus that embedded clusters more massive than about $10^{4} ~ \mathrm{M}_{\odot}$ loose a minor fraction of their stars as a result of residual gas expulsion. For less-massive clusters, such as the Orion Nebula Cluster, residual gas expulsion is more destructive with a larger expansion and leaving a smaller fraction of the initial stellar population in a bound open-cluster \citep{KAH2001}. 
Assuming an analytical time changing gas potential as used here can be consistent with simulations including heated gas, as demonstrated by \citet{Geyer2001}.

Sub-structures are examined in the studies of \citet{BanerjeeKroupa2015} or \citet{Fellhauer2009}, which indicate that the timescale for clump-merging may be shorter than the time for residual gas expulsion. As we consider high-mass clusters, the omission of sub-structures also seems reasonable.
Neglecting primordial binaries could underestimate ejections from the cluster due to the missing of very tight binaries (see also \citealt{Pfalzner2013} and references therein), again making clear that our results point to the upper limit of the bound fraction. On the other hand, primordial binaries can cause a brief initial contraction of the cluster as described in \citet{Sambaran2014} due to ``binary cooling''. Soft binaries can absorb kinetic energy to become unbound, the cooling therefor being a result of the chosen binary distribution. If the cluster re-virializes fast enough after this initial collapse, binary activity should have no major effect on the following gas expulsion phase.
The influence of $\varepsilon_\mathrm{SFE}$ varying over the cluster is related to the density profile. Should $\varepsilon_\mathrm{SFE}$ be higher in the central regions, we could expect a higher stellar density compared to a constant $\varepsilon_\mathrm{SFE}$ over the whole cluster. Simulations discussed in \citet{Pfalzner2013} compare King-type and Plummer-type clusters, showing a significant difference for the bound fraction around $\varepsilon_\mathrm{SFE} = 0.33$, whereas other $\varepsilon_\mathrm{SFE}$ has matching results. This high sensitivity to the density profile, and more generally to the phase-space density function \citep{BoilyKroupa2002,BoilyKroupa2003}, should be addressed in future studies.

Summarizing our main results:

   \begin{enumerate}
      \item The bound fraction is extremely sensitive to the timescale of gas expulsion for the considered $\varepsilon_\mathrm{SFE}$, or more precisely to the relation $\tau_\mathrm{g} / t_\mathrm{cross}$. Clusters with the same initial core density can have very different bound fractions.
      \item The variation of $\varepsilon_\mathrm{SFE}$ or the timescale of gas expulsion can result in the same bound fraction. It would be desirable to have more observational constraints, especially regarding the gas expulsion velocity.
      \item We can confirm that primordially mass segregated clusters have a smaller bound fraction. The effect lessens with increasing mass, possibly due to our models segregating during the early equilibrium phase, thus not actually depicting unsegregated heavy clusters at the beginning of gas expulsion.
      \item Stellar evolution reduces the bound fraction over the whole considered mass range. In the lower mass clusters, the first SN disturb the process of re-virialization, resulting in a more pronounced mass loss.
      \item Heating through dynamical friction could be of the same order of magnitude as the expansion caused by stellar evolution in unsegregated clusters.
      \item The massive, very compact clusters considered in our simulations seem to expand well within their tidal radii. The impact of a Galactic tidal field at solar distance is negligible, as is the more technical aspect of keeping or removing escaped stars from the simulated cluster in this case.
      \item An Arches-like cluster can survive gas expulsion in the presence of a strong Galactic tidal field.
      \item The bound fraction increases from 20\% to 80\% for $M_\mathrm{ecl} \approx 5 \times 10^{3} ~ \mathrm{M}_{\odot}$ to $10^{5} ~ \mathrm{M}_{\odot}$ for the canonical combination of gas expulsion parameters $\varepsilon_\mathrm{SFE}=0.33$, $\tau_\mathrm{d} = 0.6 ~ \mathrm{Myr}$ and $v_\mathrm{g} = 10 ~ \mathrm{km ~ s^{-1}}$.
   \end{enumerate}

\begin{acknowledgement}
 We acknowledge the support of the Argelander-Institut f\"ur Astronomie computing team.
\end{acknowledgement}

\bibliographystyle{aa} 
\bibliography{lit}

\begin{thebibliography}{54}
\expandafter\ifx\csname natexlab\endcsname\relax\def\natexlab#1{#1}\fi

\bibitem[{{Aarseth}(2003)}]{aseth2003}
{Aarseth}, S.~J. 2003, {Gravitational N-Body Simulations}, 430

\bibitem[{{Aarseth}(2012)}]{aseth2012}
{Aarseth}, S.~J. 2012, \mnras, 422, 841

\bibitem[{{Adams}(2000)}]{Adams2000}
{Adams}, F.~C. 2000, \apj, 542, 964

\bibitem[{{Adams} \& {Fatuzzo}(1996)}]{AdamsFatuzzo1996}
{Adams}, F.~C. \& {Fatuzzo}, M. 1996, \apj, 464, 256

\bibitem[{{Alves} \& {Bouy}(2012)}]{Alves2012}
{Alves}, J. \& {Bouy}, H. 2012, \aap, 547, A97

\bibitem[{{Andr{\'e}} {et~al.}(2014){Andr{\'e}}, {Di Francesco},
  {Ward-Thompson}, {Inutsuka}, {Pudritz}, \& {Pineda}}]{Andre2014}
{Andr{\'e}}, P., {Di Francesco}, J., {Ward-Thompson}, D., {et~al.} 2014,
  Protostars and Planets VI, 27

\bibitem[{{Assmann} {et~al.}(2011){Assmann}, {Fellhauer}, {Kroupa},
  {Br{\"u}ns}, \& {Smith}}]{AssmannFellhauer2011}
{Assmann}, P., {Fellhauer}, M., {Kroupa}, P., {Br{\"u}ns}, R.~C., \& {Smith},
  R. 2011, \mnras, 415, 1280

\bibitem[{{Banerjee} \& {Kroupa}(2013)}]{SambaranPavel2013}
{Banerjee}, S. \& {Kroupa}, P. 2013, \apj, 764, 29

\bibitem[{{Banerjee} \& {Kroupa}(2014)}]{Sambaran2014}
{Banerjee}, S. \& {Kroupa}, P. 2014, \apj, 787, 158

\bibitem[{{Banerjee} \& {Kroupa}(2015{\natexlab{a}})}]{2015arXiv151203074B}
{Banerjee}, S. \& {Kroupa}, P. 2015{\natexlab{a}}, ArXiv e-prints
  [\eprint[arXiv]{1512.03074}]

\bibitem[{{Banerjee} \& {Kroupa}(2015{\natexlab{b}})}]{BanerjeeKroupa2015}
{Banerjee}, S. \& {Kroupa}, P. 2015{\natexlab{b}}, \mnras, 447, 728

\bibitem[{{Bate} {et~al.}(2014){Bate}, {Tricco}, \& {Price}}]{Bate2014}
{Bate}, M.~R., {Tricco}, T.~S., \& {Price}, D.~J. 2014, \mnras, 437, 77

\bibitem[{{Baumgardt} {et~al.}(2008){Baumgardt}, {De Marchi}, \&
  {Kroupa}}]{bgetl2008}
{Baumgardt}, H., {De Marchi}, G., \& {Kroupa}, P. 2008, \apj, 685, 247

\bibitem[{{Baumgardt} \& {Kroupa}(2007)}]{Baumgardt2007}
{Baumgardt}, H. \& {Kroupa}, P. 2007, \mnras, 380, 1589

\bibitem[{{Boily} \& {Kroupa}(2002)}]{BoilyKroupa2002}
{Boily}, C. \& {Kroupa}, P. 2002, in Astronomical Society of the Pacific
  Conference Series, Vol. 285, Modes of Star Formation and the Origin of Field
  Populations, ed. E.~K. {Grebel} \& W.~{Brandner}, 141

\bibitem[{{Boily} \& {Kroupa}(2003)}]{BoilyKroupa2003}
{Boily}, C.~M. \& {Kroupa}, P. 2003, \mnras, 338, 673

\bibitem[{{Bontemps} {et~al.}(2010){Bontemps}, {Motte}, {Csengeri}, \&
  {Schneider}}]{BontempsMotte2010}
{Bontemps}, S., {Motte}, F., {Csengeri}, T., \& {Schneider}, N. 2010, \aap,
  524, A18

\bibitem[{{Clarkson} {et~al.}(2012){Clarkson}, {Ghez}, {Morris}, {Lu},
  {Stolte}, {McCrady}, {Do}, \& {Yelda}}]{Clarkson2012}
{Clarkson}, W.~I., {Ghez}, A.~M., {Morris}, M.~R., {et~al.} 2012, \apj, 751,
  132

\bibitem[{{Dale} {et~al.}(2015){Dale}, {Ercolano}, \& {Bonnell}}]{Dale2015}
{Dale}, J.~E., {Ercolano}, B., \& {Bonnell}, I.~A. 2015, ArXiv e-prints
  [\eprint[arXiv]{1504.05896}]

\bibitem[{{de Grijs} {et~al.}(2002){de Grijs}, {Gilmore}, {Johnson}, \&
  {Mackey}}]{deGrijsGilmore2002}
{de Grijs}, R., {Gilmore}, G.~F., {Johnson}, R.~A., \& {Mackey}, A.~D. 2002,
  \mnras, 331, 245

\bibitem[{{Elmegreen} \& {Elmegreen}(2006)}]{Elme2006}
{Elmegreen}, B.~G. \& {Elmegreen}, D.~M. 2006, \apj, 650, 644

\bibitem[{{Espinoza} {et~al.}(2009){Espinoza}, {Selman}, \&
  {Melnick}}]{Espinoza2009}
{Espinoza}, P., {Selman}, F.~J., \& {Melnick}, J. 2009, \aap, 501, 563

\bibitem[{{Fellhauer} {et~al.}(2009){Fellhauer}, {Wilkinson}, \&
  {Kroupa}}]{Fellhauer2009}
{Fellhauer}, M., {Wilkinson}, M.~I., \& {Kroupa}, P. 2009, \mnras, 397, 954

\bibitem[{{Figer} {et~al.}(2002){Figer}, {Najarro}, {Gilmore}, {Morris}, {Kim},
  {Serabyn}, {McLean}, {Gilbert}, {Graham}, {Larkin}, {Levenson}, \&
  {Teplitz}}]{Figer2002}
{Figer}, D.~F., {Najarro}, F., {Gilmore}, D., {et~al.} 2002, \apj, 581, 258

\bibitem[{{Geyer} \& {Burkert}(2001)}]{Geyer2001}
{Geyer}, M.~P. \& {Burkert}, A. 2001, \mnras, 323, 988

\bibitem[{{Habibi} {et~al.}(2013){Habibi}, {Stolte}, {Brandner}, {Hu{\ss}mann},
  \& {Motohara}}]{Habibi2013}
{Habibi}, M., {Stolte}, A., {Brandner}, W., {Hu{\ss}mann}, B., \& {Motohara},
  K. 2013, \aap, 556, A26

\bibitem[{{Harfst} {et~al.}(2010){Harfst}, {Portegies Zwart}, \&
  {Stolte}}]{Harfst2010}
{Harfst}, S., {Portegies Zwart}, S., \& {Stolte}, A. 2010, \mnras, 409, 628

\bibitem[{{Heggie} \& {Hut}(2003)}]{HH2003}
{Heggie}, D. \& {Hut}, P. 2003, {The Gravitational Million-Body Problem: A
  Multidisciplinary Approach to Star Cluster Dynamics}

\bibitem[{{Hurley} \& {Bekki}(2008)}]{HurleyBekki2008}
{Hurley}, J.~R. \& {Bekki}, K. 2008, \mnras, 389, L61

\bibitem[{{Hurley} {et~al.}(2000){Hurley}, {Pols}, \& {Tout}}]{hur2000}
{Hurley}, J.~R., {Pols}, O.~R., \& {Tout}, C.~A. 2000, \mnras, 315, 543

\bibitem[{{Hurley} {et~al.}(2002){Hurley}, {Tout}, \& {Pols}}]{hur2002}
{Hurley}, J.~R., {Tout}, C.~A., \& {Pols}, O.~R. 2002, \mnras, 329, 897

\bibitem[{{Kim} {et~al.}(2000){Kim}, {Figer}, {Lee}, \& {Morris}}]{Kim2000}
{Kim}, S.~S., {Figer}, D.~F., {Lee}, H.~M., \& {Morris}, M. 2000, \apj, 545,
  301

\bibitem[{{Kroupa}(2002)}]{PavelDisc}
{Kroupa}, P. 2002, \mnras, 330, 707

\bibitem[{{Kroupa}(2005)}]{Kroupa2005}
{Kroupa}, P. 2005, in ESA Special Publication, Vol. 576, The Three-Dimensional
  Universe with Gaia, ed. C.~{Turon}, K.~S. {O'Flaherty}, \& M.~A.~C.
  {Perryman}, 629

\bibitem[{{Kroupa}(2008)}]{Kroupa2008initial}
{Kroupa}, P. 2008, in Lecture Notes in Physics, Berlin Springer Verlag, Vol.
  760, The Cambridge N-Body Lectures, ed. S.~J. {Aarseth}, C.~A. {Tout}, \&
  R.~A. {Mardling}, 181

\bibitem[{{Kroupa} {et~al.}(2001){Kroupa}, {Aarseth}, \& {Hurley}}]{KAH2001}
{Kroupa}, P., {Aarseth}, S., \& {Hurley}, J. 2001, \mnras, 321, 699

\bibitem[{{Kroupa} {et~al.}(2013){Kroupa}, {Weidner}, {Pflamm-Altenburg},
  {Thies}, {Dabringhausen}, {Marks}, \& {Maschberger}}]{KroupaWeidner2013}
{Kroupa}, P., {Weidner}, C., {Pflamm-Altenburg}, J., {et~al.} 2013, {The
  Stellar and Sub-Stellar Initial Mass Function of Simple and Composite
  Populations}, ed. T.~D. {Oswalt} \& G.~{Gilmore}, 115

\bibitem[{{Krumholz} \& {Matzner}(2009)}]{KrumholzMatzner2009}
{Krumholz}, M.~R. \& {Matzner}, C.~D. 2009, \apj, 703, 1352

\bibitem[{{Lada} \& {Lada}(2003)}]{LadaLada2003}
{Lada}, C.~J. \& {Lada}, E.~A. 2003, \araa, 41, 57

\bibitem[{{Lada} {et~al.}(1984){Lada}, {Margulis}, \& {Dearborn}}]{Lada1984}
{Lada}, C.~J., {Margulis}, M., \& {Dearborn}, D. 1984, \apj, 285, 141

\bibitem[{{Launhardt} {et~al.}(2002){Launhardt}, {Zylka}, \&
  {Mezger}}]{Launhardt2002}
{Launhardt}, R., {Zylka}, R., \& {Mezger}, P.~G. 2002, \aap, 384, 112

\bibitem[{{Littlefair} {et~al.}(2003){Littlefair}, {Naylor}, {Jeffries},
  {Devey}, \& {Vine}}]{LittlefairNaylor2003}
{Littlefair}, S.~P., {Naylor}, T., {Jeffries}, R.~D., {Devey}, C.~R., \&
  {Vine}, S. 2003, \mnras, 345, 1205

\bibitem[{{Machida} \& {Matsumoto}(2012)}]{MM2012}
{Machida}, M.~N. \& {Matsumoto}, T. 2012, \mnras, 421, 588

\bibitem[{{Malinen} {et~al.}(2012){Malinen}, {Juvela}, {Rawlings},
  {Ward-Thompson}, {Palmeirim}, \& {Andr{\'e}}}]{Malinen2012}
{Malinen}, J., {Juvela}, M., {Rawlings}, M.~G., {et~al.} 2012, \aap, 544, A50

\bibitem[{{McMillan} {et~al.}(2007){McMillan}, {Vesperini}, \& {Portegies
  Zwart}}]{McMillanVesperini2007}
{McMillan}, S.~L.~W., {Vesperini}, E., \& {Portegies Zwart}, S.~F. 2007, \apjl,
  655, L45

\bibitem[{{Megeath} {et~al.}(2016){Megeath}, {Gutermuth}, {Muzerolle},
  {Kryukova}, {Hora}, {Allen}, {Flaherty}, {Hartmann}, {Myers}, {Pipher},
  {Stauffer}, {Young}, \& {Fazio}}]{Megeath2016}
{Megeath}, S.~T., {Gutermuth}, R., {Muzerolle}, J., {et~al.} 2016, \aj, 151, 5

\bibitem[{{Meylan}(2000)}]{Meylan2000}
{Meylan}, G. 2000, in Astronomical Society of the Pacific Conference Series,
  Vol. 211, Massive Stellar Clusters, ed. A.~{Lan{\c c}on} \& C.~M. {Boily},
  215

\bibitem[{{Nagata} {et~al.}(1995){Nagata}, {Woodward}, {Shure}, \&
  {Kobayashi}}]{Nagata1995}
{Nagata}, T., {Woodward}, C.~E., {Shure}, M., \& {Kobayashi}, N. 1995, \aj,
  109, 1676

\bibitem[{{Olczak} {et~al.}(2012){Olczak}, {Kaczmarek}, {Harfst}, {Pfalzner},
  \& {Portegies Zwart}}]{Olczak2012}
{Olczak}, C., {Kaczmarek}, T., {Harfst}, S., {Pfalzner}, S., \& {Portegies
  Zwart}, S. 2012, \apj, 756, 123

\bibitem[{{Pfalzner} \& {Kaczmarek}(2013)}]{Pfalzner2013}
{Pfalzner}, S. \& {Kaczmarek}, T. 2013, \aap, 555, A135

\bibitem[{{Plummer}(1911)}]{Plummer1911}
{Plummer}, H.~C. 1911, \mnras, 71, 460

\bibitem[{{Smith} {et~al.}(2013){Smith}, {Goodwin}, {Fellhauer}, \&
  {Assmann}}]{SmithGoodwin2013}
{Smith}, R., {Goodwin}, S., {Fellhauer}, M., \& {Assmann}, P. 2013, \mnras,
  428, 1303

\bibitem[{{Stolte} {et~al.}(2008){Stolte}, {Ghez}, {Morris}, {Lu}, {Brandner},
  \& {Matthews}}]{Stolte2008}
{Stolte}, A., {Ghez}, A.~M., {Morris}, M., {et~al.} 2008, \apj, 675, 1278

\bibitem[{{Wuchterl} \& {Tscharnuter}(2003)}]{WuchterlTscharnuter2003}
{Wuchterl}, G. \& {Tscharnuter}, W.~M. 2003, \aap, 398, 1081

\end{thebibliography}

\end{document}